\documentclass[10pt]{article}
\usepackage{epsfig}
\usepackage{amsfonts}

\newcommand{\nwc}{\newcommand}
\nwc{\R}{{\mathbb R}}
\nwc{\C}{{\mathbb C}}
\nwc{\Z}{{\mathbb Z}}
\nwc{\N}{{\mathbb N}}

\newtheorem{theorem}{Theorem}[section]
\newtheorem{lemma}[theorem]{Lemma}
\newtheorem{remark}[theorem]{Remark}

\newtheorem{proposition}[theorem]{Proposition}
\newtheorem{corollary}[theorem]{Corollary}

\nwc{\address}{}
\nwc{\ams}[1]{}
\nwc{\submitted}{}
\nwc{\proof}{\smallskip\noindent{\bf Proof } }
\nwc{\ftnote}[1]{\footnote{#1}}

\nwc{\bm}{\boldmath}
\nwc{\ubm}{\unboldmath} 
\nwc{\m}{\mbox}
\nwc{\ds}{\displaystyle} 
\nwc{\reseteqnos}{
 \setcounter{equation}{0}  }

\nwc{\ip}[1]{{\langle #1 \rangle}}
\nwc{\ipbig}[1]{{\left\langle #1 \right\rangle}}
\nwc{\Span}[1]{\mathop{\rm span}\nolimits\{#1\}}
\nwc{\sfrac}[2]{{\textstyle\frac{ #1}{ #2}}}

\def\drawbox#1#2{
   \hbox{\vrule\vbox{\hrule width#1\vskip #2\hrule width#1}\vrule}}
\nwc{\eproof}{\drawbox{.1in}{.1in}\smallskip}

\nwc{\hide}[1]{#1}     
\nwc{\eq}[1]{(\ref{#1})}
\nwc{\figref}[1]{\ref{#1}} 

\nwc{\be}{\begin{equation}}
\nwc{\ee}{\end{equation}}
\nwc{\bee}{\begin{eqnarray}}
\nwc{\eee}{\end{eqnarray}}
\nwc{\ba}{\begin{array}}
\nwc{\ea}{\end{array}}
\nwc{\benum}{\begin{enumerate}}
\nwc{\eenum}{\end{enumerate}}
\nwc{\posthat}{\hat{\phantom{l}}}
\nwc{\naive}{na{\"\i}ve}

\renewcommand{\Im}{\mathop{\rm Im}\nolimits}
\renewcommand{\Re}{\mathop{\rm Re}\nolimits}
\nwc{\Res}{\mathop{\rm Res}}
\nwc{\half}{\frac12}
\nwc{\third}{\frac13}
\nwc{\quarter}{\frac14}
\nwc{\shalf}{{\textstyle\frac12}}
\nwc{\ep}{\varepsilon}
\nwc{\eps}{\varepsilon}
\nwc{\eptozero}{\quad\mbox{ as\quad$\varepsilon\to0$}}

\nwc{\diag}{\mathop{\rm diag}\nolimits}
\nwc{\sech}{\mathop{\rm sech}\nolimits}
\nwc{\D}{\partial}
\nwc{\intR}{\int_{-\infty}^\infty}
\nwc{\sumZ}{\sum_{j=-\infty}^\infty }

\nwc{\spec}{\sigma}
\nwc{\specess}{\spec_{\rm ess}}
\nwc{\specdisc}{\spec_{\rm disc}}

\nwc{\calA}{{\cal A}}
\nwc{\ws}{w}
\nwc{\bmy}{\m{\bm $y$\ubm}}
\nwc{\damp}{d_0}
\nwc{\eamp}{d_\infty}
\nwc{\tdamp}{\tilde{d}_0}
\nwc{\teamp}{\tilde{d}_\infty}
\nwc{\wmax}{w_{\rm max}}
\nwc{\wmaxs}{a_*} 
\nwc{\phis}{\phi_*}
\nwc{\fs}{f_*}
\nwc{\Fs}{F_*}
\nwc{\omegas}{\omega_*}
\nwc{\Fmax}{F_{\rm max}}
\nwc{\FA}{F_A}
\nwc{\yf}{\bmy}
\nwc{\Bf}{\bmB}
\nwc{\tBf}{\m{\bm $\tilde{B}$\ubm}}
\nwc{\jpm}{{(j+m)}}
\nwc{\jmm}{{(j-m)}}
\nwc{\jmax}{j_{\rm max}}
\nwc{\Lmax}{\Lambda_{\rm max}}
\nwc{\mum}{\omega_{\rm cr}}
\nwc{\Rin}{R_{\rm in}}
\nwc{\Rout}{R_{\rm out}}

\nwc{\bfy}{{\bf y}}
\nwc{\bfz}{{\bf z}}

\nwc{\pdis}{{\cal P}}
\nwc{\lamcr}{\lambda_{\rm cr}}

\nwc{\bmu}{\m{\bm$u$\ubm}} 
\nwc{\bmB}{\m{\bm $B$\ubm}}

\nwc{\todo}[1]{\medskip\noindent{\bf TODO:}#1 \medskip}
\nwc{\query}[1]{\medskip\noindent{\bf QUERY:}#1 \medskip}
\nwc{\topic}[1]{\smallskip{\em #1}\smallskip}

\nwc{\halo}[1]{\mathop {#1}\limits^\circ}
\nwc{\Ccut}{\C_\sigma}
\nwc{\bmzeta}{\m{\bm $\zeta$\ubm}}
\nwc{\zetaf}{\bmzeta}
\nwc{\tzetaf}{\BM{\tilde{\zeta}}}
\nwc{\bmC}{\m{\bm $C$\ubm}}
\nwc{\Cf}{\bmC}
\nwc{\tCf}{\m{\bm $\tilde{C}$\ubm}}
\nwc{\bma}{\m{\bm $a$\ubm}}
\nwc{\af}{\bma}
\nwc{\hyf}{{\halo{\yf}}}
\nwc{\trace}{{\mathop{\rm tr}\nolimits}}
\nwc{\hY}{{\halo{\m{\bm $Y$\ubm}}}}

\nwc{\BM}[1]{\m{\bm $#1$\ubm}}
\nwc{\bmb}{\BM{b}}
\nwc{\bme}{\BM{e}}
\nwc{\tyf}{\BM{\tilde{y}}}
\nwc{\ysix}{\hat{\BM{y}}}
\nwc{\zf}{\BM{z}}
\nwc{\tzf}{\BM{\tilde{z}}}
\nwc{\zsix}{\hat{\BM{z}}}
\nwc{\Bfour}{B}
\nwc{\Bsix}{B_{\wedge}} 
\nwc{\Bfsix}{\Bf_{\wedge}}
\nwc{\ywedgety}{\yf\wedge\tyf}
\nwc{\zwedgetz}{\zf\wedge\tzf}
\nwc{\ylwyl}{\yf^\lt_1\wedge\yf^\lt_2}
\nwc{\zrwzr}{\zf^\rt_1\wedge\zf^\rt_2}
\nwc{\hylwyl}{\hyf_1\wedge\hyf_2}
\nwc{\hzrwzr}{\hzf_1\wedge\hzf_2}
\nwc{\hatbmb}{\BM{\hat{b}}}
\nwc{\haloB}{\halo{\Bf}}
\nwc{\halobz}{\BM{\halo{z}}}
\nwc{\hzf}{{\halo{\zf}}}
\nwc{\hZ}{{\halo{\m{\bm $Z$\ubm}}}}
\nwc{\rt}{{(\infty)}}
\nwc{\lt}{{(0)}}
\nwc{\li}{{(-\infty)}}
\nwc{\zzyy}[2]{\zf^\rt_{#1}\cdot\yf^\lt_{#2}}
\nwc{\Evans}{E}
\nwc{\evpj}{\eq{ysys}--\eq{ysysbc}\ }
\nwc{\bmeta}{\m{\bm $\eta$\ubm}}
\nwc{\etaf}{\bmeta}
\nwc{\Msix}{\hat{M}}
\nwc{\bmm}{\BM{m}}
\nwc{\bmn}{\BM{n}}
\nwc{\bmms}{\hat{\BM{m}}}
\nwc{\bmns}{\hat{\BM{n}}}
\nwc{\bmY}{\m{\bm $Y$\ubm}}
\nwc{\bmD}{\m{\bm $D$\ubm}}
\nwc{\bmK}{\m{\bm $K$\ubm}}
\nwc{\bmv}{\m{\bm $v$\ubm}}
\nwc{\bmM}{\m{\bm $M$\ubm}}

\nwc{\alphadet}{\alpha_{11}\alpha_{22}-\alpha_{21}\alpha_{12}}
\nwc{\Cfsix}{\Cf_{\wedge}}
\nwc{\zetas}{\hat{\zetaf}}
\nwc{\etas}{\hat{\etaf}}
\nwc{\musix}{\hat{\mu}}
\nwc{\vsixlt}{\hat{\BM{v}}^\li}
\nwc{\wsixrt}{\hat{\BM{w}}^\rt}
\nwc{\vz}{\BM{v}}
\nwc{\tvz}{\BM{\tilde{v}}}
\nwc{\calF}{{\cal F}}

\newcommand{\ccite}[1]{\cite{#1}}  

\begin{document}
\pagestyle{myheadings}

\title{
Spectrally stable encapsulated vortices \\
for nonlinear Schr\"odinger equations}

\author{R. L. Pego\thanks{
Department of Mathematics \& 
Institute for Physical Science and Technology, 
University of Maryland, 
College Park MD 20742}
\quad and \  
H. A. Warchall\thanks{ 
Department of Mathematics, 
University of North Texas, 
Denton TX 76203 and Division of Mathematical Sciences,
National Science Foundation}
}

\date{June 2001 (revised January 2002)}
\maketitle

\begin{abstract}

A large class of multidimensional nonlinear Schr\"odinger equations admit
localized nonradial standing wave solutions 
that carry nonzero intrinsic angular momentum.  
Here we provide evidence that certain of these spinning excitations are 
spectrally stable.  We find such waves for equations 
in two space dimensions with {\it focusing-defocusing} nonlinearities, such as
cubic-quintic.
Spectrally stable waves resemble a vortex (non-localized solution with
asymptotically constant amplitude) cut off at large radius by a
kink layer that exponentially localizes the solution.

For the evolution equations linearized about a localized spinning wave, 
we prove that unstable eigenvalues are zeros of 
Evans functions for a finite set of ordinary differential equations.
Numerical computations indicate that
there exist spectrally stable standing waves having central vortex of any degree.
\end{abstract}

\vskip .1 in
\noindent
{\bf Running head:} Stability of encapsulated vortices

\newcommand{\inn}{\input} 
\section{Introduction}
\reseteqnos

A long-term goal of the study of nonlinear waves is to sort out
what kinds of structures and interactions are significant
and robust. From this point of view, nonlinear Schr\"odinger equations
are worthy of study because of their status as a
canonical model for weakly nonlinear phenomena in many fields 
(including nonlinear optics as a prominent example),
and as a simple model for more complicated gauge theories of 
mathematical physics. In more than one space dimension 
such equations admit a variety of solutions with interesting structure,
including solitary waves (e.g., ``light bullets'') and vortices
(kin to the magnetic vortices of the theory of superconductivity).

Here we consider nonlinear Schr\"odinger equations of the form
\be
-i \,\partial_t u - \Delta u = g(u)
\label{1nls}
\ee
where $u{\colon}\,{\R}^{2+1}\to {\C}$, and where 
$g(u)=h(|u|^2)u$ for some real-valued $C^1$ function $h$.
Our most significant results concern nonlinearities for which
focusing ($h'>0$) at small amplitude  is overcome by defocusing
($h'<0$) at larger amplitude.
Prototypical is the cubic-quintic nonlinearity
\be
\label{1g35}
g(u) = |u|^2u-|u|^4u.
\ee
Much of the theory developed in this paper also applies to 
equations of focusing type.

We are interested in the stability of solitary-wave solutions 
with the symmetry of a vortex, having the special form 
\be
u_0(x,t) =  e^{i\omega t+im\theta }\ws(r)
\label{1sw}
\ee
in terms of polar coordinates $(r,\theta)$ for ${\R}^2$. 
Here $\omega$ is a real constant (the standing wave frequency), 
$m$ is an integer (the vortex degree), and $\ws{\colon}\,[0,\infty
)\to {\R}$. 
We also refer to $m$ as the spin of the wave, because
the conserved quantity 
$L$
that corresponds via Noether's principle to the
rotational invariance of \eq{1nls} is the 
(third component of the)
angular momentum
\[
\int_{\R^2} (-i x\times\nabla u)\overline{u} .
\]
For standing waves having the form \eq{1sw}, $L=2\pi m\int_0^\infty
w^2r\,dr$.

The standing-wave amplitude profile $\ws$ in \eq{1sw} must satisfy
the differential equation
\be
\ws''+{{1 \over r}}\ws'-{{{m^2} \over {r^2}}}\ws
+f(\ws)=0,
\label{1weq}
\ee
where $f(w)=g(w)-\omega w$.
It is known that, under suitable conditions on $f$, for each integer $m$
this equation has $C^2$ solutions $w$ such that
$w(r)\to0$ exponentially as $r\to\infty$. 
Iaia and Warchall \ccite{IW3} proved that 
for fixed $m$, there is at least one such solution $w$ for each prescribed
number $n$ of positive zeros.
Each such solution gives rise to a localized standing
wave solution of (\ref{1nls}).  
In turn, this gives rise to solitary-wave solutions traveling at any
speed, via the Galilean boost transformation for the
Schr\"odinger equation.

Without loss of generality we assume $h(0)=0$.
Then the restriction of $g$ to the real axis is a $C^1$ function
with $g'(0) = 0$ and $g'(s)=O(s^2)$ as $s\to0$.
For an exponentially localized solution to exist, it is necessary that
$f'(0)<0$, so the standing-wave frequency $\omega$ is positive.
It is also necessary that
$F(s)\equiv\int_0^s f(\sigma)\,d\sigma$ be positive 
for some positive value of $s$ attained by profile amplitudes $w(r)$.  

In this paper we investigate the linear stability of such exponentially 
localized
solitary waves. Linearizing (\ref{1nls}) about such a solution yields
a real system of the form $\D_t{\Phi}=A{\Phi}$ where $A$ is a
$2\times2$ matrix of second-order partial differential operators that has
purely imaginary essential spectrum 
$\specess=\{i\tau: |\tau|\ge\omega\}$.
 Any eigenvalues of $A$ not in the
essential spectrum are discrete
(isolated and of finite multiplicity) and occur with Hamiltonian
symmetry: If $\lambda$ is an eigenvalue, then so are $-\lambda$ and
$\overline{\lambda}$. 

We prove that the search for unstable eigenvalues of $A$ reduces to 
the study of a finite number of eigenvalue problems for
$2\times2$ systems of second-order ordinary differential equations.
Eigenvalues for each of these systems correspond to zeros of an Evans
function or Wronskian that we prove is a globally analytic function in 
the cut plane $\Ccut=\C\setminus\specess$.
The real part of any unstable eigenvalue is bounded by a quantity
determined from the profile shape.

We use the argument principle and numerical computation to locate
zeros of the Evans functions. 
Very often the operator $A$ has unstable eigenvalues
(meaning eigenvalues with positive real part), indicating that the
associated solitary wave is linearly exponentially unstable with
respect to perturbations of initial data.

However, we have discovered spectrally stable waves (waves with no unstable
eigenvalues) in a certain parameter regime for several nonlinearities
of focusing-defocusing type, including the cubic-quintic 
\eq{1g35}.  All of these waves have
radial profiles $w(r)$ with no positive zeros.
We have found such waves for all values of spin index satisfying $|m|\le 5$,
with indications that they exist for arbitrary $m$. 

The spectrally stable spinning waves ($m\ne0$)
are found when $\omega$ is close to but less than a critical value
$\omegas$ for which the two bounded regions between 
the graphs of $y=g(x)$ and $y=\omegas x$ for $x\ge0$ have 
{\it equal area}.  
This condition on $\omegas$ is equivalent to the condition that
the potential function $F$ with $\omega=\omegas$  has a zero-height 
local maximum or ``hilltop'' at a point $\wmaxs>0$.
For the cubic-quintic case $\omegas=\frac{3}{16}=0.1875$ and
spectrally stable waves are found for $\omega$ in an interval
$\mum(m)<\omega<\omegas$ 
which shrinks as $|m|$ increases,
with width roughly proportional to $|m|^{-2}$. 

A transition to instability occurs
as the standing wave frequency $\omega$ decreases below $\mum$.
Unstable eigenvalues appear, for a particular azimuthal mode,
due to collisions of pairs of imaginary 
eigenvalues as in a Hamiltonian Hopf bifurcation. 

The waves develop a distinctive structure  
as $\omega$ approaches $\omegas$ from below. 
(See Figs.~\figref{FCQVaryOmegam1} and \figref{FCQVaryOmegam3} in section 2.)
The wave amplitude $w(r)$ increases from zero to remain
approximately constant over a large $r$-interval,
taking values near $\wmaxs$, 
near the hilltop of the nonlinear potential $F$, before
rapidly returning exponentially to zero. Thus the waves have a
structure consisting of a core region resembling a {\it vortex} (a
non-localized solution with asymptotically constant amplitude, 
{cf.} \ccite{Neu}), 
{encapsulated} or cut off (at an apparently arbitrarily large radius) by a
circular {\it kink} region that serves to exponentially localize the solution.  

What are the salient features of the nonlinearity that might explain 
the existence of stable encapsulated vortices with this kind of 
structure? We believe that a basic
understanding begins with the stability of spatially constant solutions
\be \label{1const}
u(x,t) = a e^{i\omega t}
\ee
where $a>0$ is constant and $\omega=g(a)/a$. 
This solution is linearly unstable if $h'(a^2)>0$ (the focusing case) and is
linearly stable if $h'(a^2)<0$ (defocusing); see \cite{NK} and
Remark~\ref{const-stab} below. 
Equivalently, stability is determined by the sign of $g'(a)-g(a)/a$,
a quantity that is easily visualized on the graph of $g$ 
over the positive reals.
The spatially constant solution $ae^{i\omega t}$ is linearly 
unstable if $g'(a)-g(a)/a>0$ and linearly stable if $g'(a)-g(a)/a<0$.

\begin{figure} 
\caption{Equal area construction for kink frequency}
\label{Equal}
\begin{center} \leavevmode { \hbox{
        \epsfxsize=\xsz
        \epsfysize=\ysz
        \epsffile{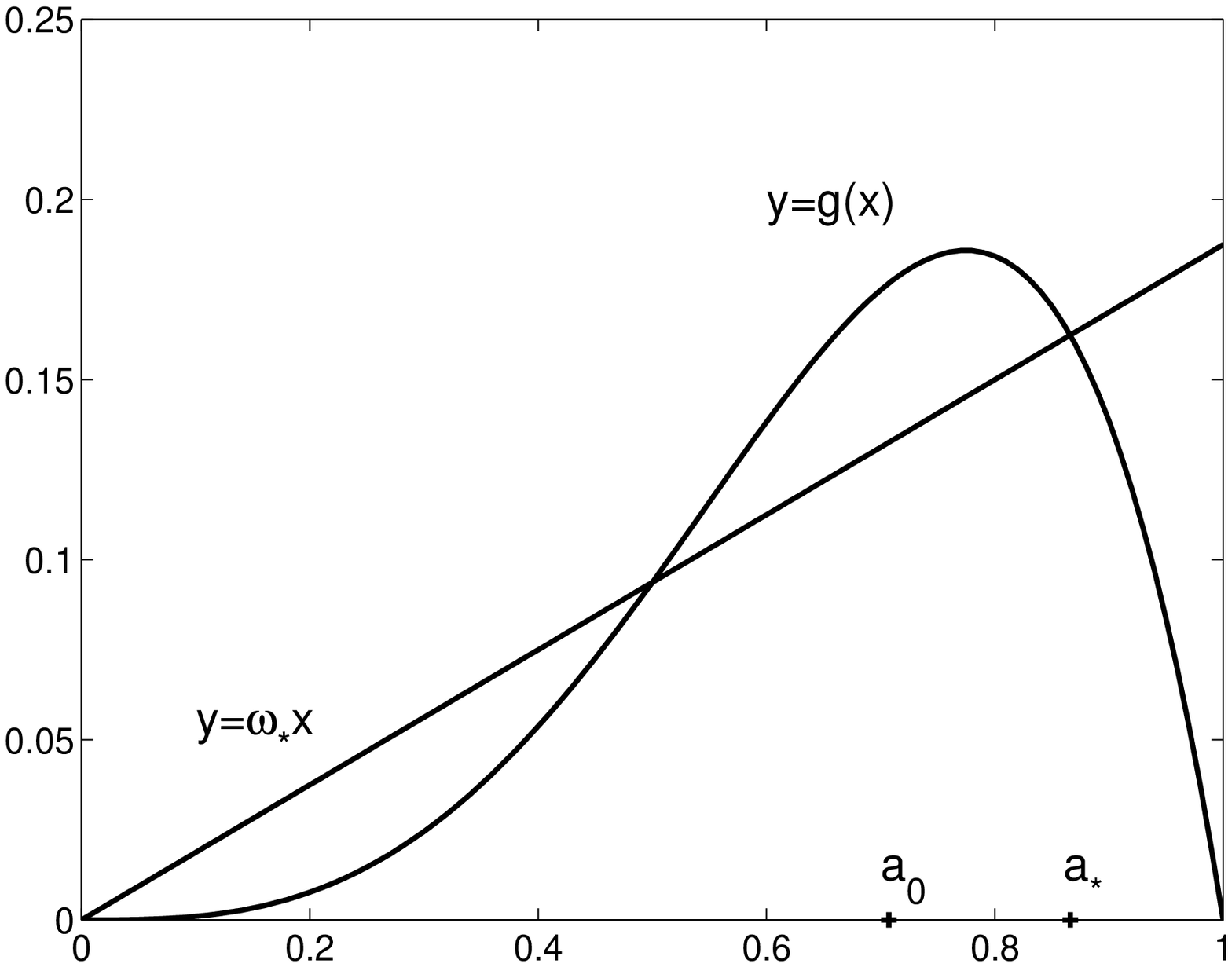}
} } \end{center} 
\end{figure}

A focusing-defocusing nonlinearity like
the cubic-quintic \eq{1g35} has two features that appear to be key
(see Fig.~\figref{Equal}):
\begin{itemize}
\item[(i)] There is a number $a_0>0$ such that 
{the constant-state solution \eq{1const} is linearly unstable for
$0<a<a_0$ ($h'(a^2)>0$) and is linearly stable for $a>a_0$ ($h'(a^2)<0$).}
\item[(ii)]
There is a number
$\omegas>0$ for which the graphs of $y=g(x)$ and $y=\omegas x$ 
enclose two regions of equal area, and so for which the potential $F$
has zero-height hilltops at $s=0$ and at $s=\wmaxs>0$. 
\end{itemize}
As we discuss in
section 6, for any nonlinearity with these properties 
there is a linearly stable one-dimensional kink
solution of \eq{1nls} that connects the state
$\wmaxs e^{i\omegas t}$ to the zero state. 
The value $\omegas$ is the standing-wave frequency of the kink.
In the limit as $\omega$ approaches $\omegas$ and $|m|\to\infty$,
we will see that the structure of the encapsulated vortices consists 
of parts that approximate linearly stable solutions (constant states or kinks). 
In this way it becomes plausible that these waves are linearly stable.

The nonlinear Schr\"odinger equation is of fundamental importance in 
nonlinear optics, where a cubic (Kerr) nonlinearity is most
often assumed. A substantial body of work exists on so-called
saturable media, for which the nonlinearity takes the form
\be\label{1gsat}
g(u) = \frac{|u|^2u}{1+|u|^2}.
\ee
The cubic-quintic nonlinearity \eq{1g35} can be obtained by 
truncating the Taylor expansion
for this nonlinearity, but in fact there is a substantial difference
between the saturable and cubic-quintic nonlinearities at large amplitude. 
This difference is reflected in the stability of spinning solitary waves.  
In the saturable case the spinning waves with $|m|=1$ have been found 
to be unstable \ccite{FS}.   
Some spinning waves with $|m|=1$ for the cubic-quintic
nonlinearity \eq{1g35}, however, have previously been found to be stable in
direct numerical simulations by Quiroga-Teixeiro and Michinel \ccite{QM}.
We conjecture that the crucial
difference is that the cubic-quintic is focusing-defocusing, but the saturable
nonlinearity is only focusing --- it does not admit any linearly stable 
nonzero constant states.

There is experimental evidence that the cubic-quintic model is in fact 
appropriate for certain optical materials \ccite{LCTS,LCKT,LS,FSWZ}.
Instabilities of three-dimensional spinning waves for the cubic-quintic
model have been investigated numerically in \cite{DMM, MM1, MM2}.
After this paper was first submitted, we learned of work of Towers 
{\it et al.}\ on the two-dimensional cubic-quintic model \cite{TB2}. 
These authors numerically obtain stability results essentially 
similar to ours, with one discrepancy concerning the existence of
a very weak instability for a range of values 
of $\omega$ in cases for $|m|=1, 2, 3$ where we find stability.
For related work on solitary waves with vortex symmetry
in nonlinear Ginzburg-Landau equations and quadratically nonlinear 
optical media, see \cite{CM, TB1}.

The cubic-quintic nonlinear Schr\"odinger equation can also been 
considered a nonrelativistic version of a simple model in nuclear physics 
that admits nontopological solitons called ``$Q$-balls'' \cite{Coleman}.
The results of this paper may be relevant to the issue of the stability
of closed cylindrical $Q$-walls raised in \cite{AFKP}.

For radially symmetric waves ($m=0$), an energy-based criterion for
instability was established by Shatah and Strauss \ccite{SS} 
and a stability criterion was established by Grillakis, Shatah and
Strauss \ccite{GSS2}.  The result is that waves in a family parametrized by
$\omega$ are stable if $dN/d\omega>0$ and are unstable if $dN/d\omega<0$,
where $N=\int_{\R^2}|u_0|^2$.
Consistent with previous studies \ccite{DeA,DRSOA,WLTS}
we find that
all spinless waves for the cubic-quintic \eq{1g35} are stable, but if
the sign of the quintic term is reversed then all waves become unstable.
Our numerical results elucidate the mechanism of the transition to
instability as the coefficient of the quintic term changes sign
continuously.  A single pair of imaginary eigenvalues collides at the
origin and a pair of real eigenvalues emerge with opposite sign.  
A related scenario 
occurs for solitary waves of generalized Korteweg-de Vries equations \ccite{PW92} 
in which unstable eigenvalues emerge at the origin (out of the 
continuous spectrum) as the sign of $dN/d\omega$ changes,
and suggests that stability criteria based on the sign of $dN/d\omega$
will not detect instability transitions that are not associated with
eigenvalues moving through the origin.

For the spinning solitary waves of the nonlinear Schr\"odinger equation under
consideration ($m\ne0$), it is not known whether spectral stability
is mathematically a sufficient condition for nonlinear stability 
(modulo symmetries of the wave family as appropriate). 
It is conceivable
that instability could be created by nonlinear resonance phenomena
involving both localized oscillatory modes and radiation modes in the
continuous spectrum, such as have been studied by Soffer and Weinstein
\ccite{SW99}.
If these waves ultimately turn out to
be unstable, however, the absence of unstable eigenvalues suggests that
they would enjoy a long ``lifetime'' under perturbation, and that any
instability mechanism would be a subtle one.

This paper is organized as follows. 
In section 2, we survey the structure of solitary wave
solutions to \eq{1nls} as the standing-wave frequency, spin, and node
number vary, contrasting the cubic-quintic nonlinearity \eq{1g35} with
a pure cubic.
We linearize about a solitary wave and prove some fundamental results
concerning the spectrum of the operator $A$ in section 3. 
In particular, in section 3.2 we establish that we can locate all unstable eigenvalues of $A$ by analyzing a finite number of ordinary differential equations. 
In section 4 we introduce and analyze the Evans functions and 
Wronskians associated with the eigenvalue problem. 
A representation of the Evans function in terms of exterior products, 
described by Alexander and Sachs \ccite{AS}, 
proves to be convenient for purposes both theoretical and numerical.
We present numerical results in section 5. 
In section 6 we discuss heuristic reasons 
why encapsulated-vortex solutions may become stable in a limiting
parameter regime.

\section{Taxonomy of solitary waves with spin}\label{tax}
\reseteqnos

In this section we illustrate the structure of the solitary-wave
profiles $w(r)$ for various choices of spin $m$, node number $n$ and
standing wave frequency $\omega$, contrasting two different types
of nonlinearity for which existence proofs of infinite families of
localized standing waves have been given \ccite{IW1,IW3,IWW}.
In particular we contrast the cubic-quintic nonlinearity in \eq{1g35}
with the pure cubic
\be \label{2g3}
g(u) = |u|^2u.
\ee
Because of the invariance of \eq{1nls} under spatial reflection, we
may assume that $m\ge0$. 

Localized solutions of equation (\ref{1weq}) for standing wave
profiles have the asymptotic behavior (see ~\ccite{IW1})
\be
w(r) \sim \cases{ \damp r^m & as $r\to0^+,$ \cr
 \ds\frac{\eamp e^{-\sigma r}}{{\sqrt r}} & as $r\to\infty$,  }
\label{2asymp}
\ee
for some constants $\damp$ and $\eamp$, where 
$\sigma=\sqrt{-f'(0)}=\sqrt\omega>0$.
More precisely, where its amplitude is small the behavior of $w(r)$ is
governed by the linear equation
\be
w''+\frac1r w' -\frac{m^2}{r^2} w - \sigma^2 w=0. 
\label{2lin1}
\ee
Linearly independent solutions of this equation are the modified
Bessel functions $I_m(\sigma r)$ and $K_m(\sigma r)$, which have the
asymptotic behavior
\be
\ba{lll}
\ds I_m(\sigma r) \sim \frac{1}{2^m m!}(\sigma r)^m, &\quad
\ds K_m(\sigma r) \sim 2^{m-1}(m-1)!\, (\sigma r)^{-m}, &\quad
(r\to 0^+) \cr
\ds I_m(\sigma r) \sim \frac{1}{\sqrt{2\pi\sigma r}}\,e^{\sigma r}, &\quad
\ds K_m(\sigma r) \sim \sqrt{\frac{\pi}{2\sigma r}}\, e^{-\sigma r} &\quad
(r\to\infty)
\ea
\label{2IK}
\ee
except in the case $m=0$, when $K_0(\sigma r)\sim -\ln r$ as $r\to0^+$.

We compute solutions to equation (\ref{1weq}) numerically
by a shooting method, based on small-amplitude approximations of the form
$w(r)\sim \tdamp I_m(\sigma r)$ for small $r$ ($m>0$),
$w(r)\sim\teamp K_m(\sigma r)$ for large $r$.  
For a suitably small initial radius $r_0$, we numerically solve the 
initial-value problem for \eq{1weq} with initial data 
$w(r_0)=\tdamp I_m(\sigma r_0)$ and  
$w'(r_0)=\tdamp \sigma {I'}_m(\sigma r_0)$.
We vary the parameter $\tdamp $ (equivalently
$\damp$) so that at some large radius $r_1$, the solution and its
derivative are small and assume the proper asymptotic form, satisfying
$w'(r_1)/w(r_1)=\sigma {K'}_m(\sigma r_1)/K_m(\sigma r_1)$.

\begin{figure} 
\caption{Potentials for cubic and cubic-quintic nonlinearities}
\label{Fpot}
\begin{center} \leavevmode { \hbox{
        \epsfxsize=\xsz
        \epsfysize=\ysz
        \epsffile{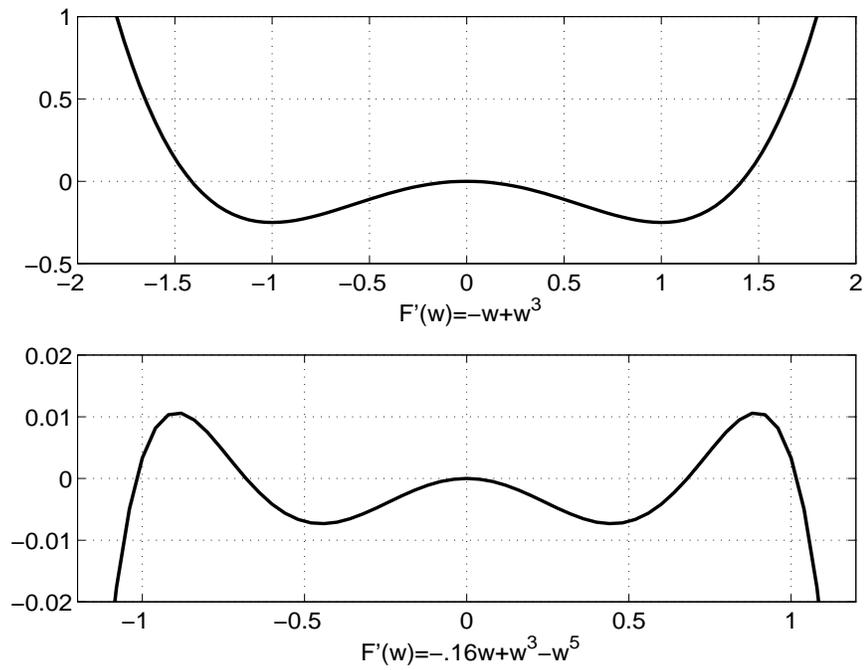}
} } \end{center} 
\end{figure}

In Fig.~\figref{Fpot} we plot the potential $F$ for the cubic and cubic-quintic
nonlinearities, taking $\omega=1$ and $\omega=.175$ respectively.
We may interpret (\ref{1weq}) as an equation of motion for a point
with position $w(r)$ 
at time $r$
moving in the potential well
$F(w)-\half{m^2}w^2/{r^2}$, subject to the time-dependent damping
force $-w'/r$. 
As $r\to\infty$, the particle ``balances'' on top of the hill at $w=0$.
Nodes in the radial profile occur when the particle crosses the top of
this hill. Since the hilltop
at the origin is an unstable equilibrium of this one-dimensional
``motion,'' solutions satisfying $w(r)\to0$ as $r\to\infty$ 
are found for $\tdamp $ in a discrete set.

\begin{figure} 
\caption{Nodeless profiles for cubic with varying spin $m$}
\label{FCubicVarySpin}
\begin{center} \leavevmode { \hbox{
        \epsfxsize=\xsz
        \epsfysize=\ysz
        \epsffile{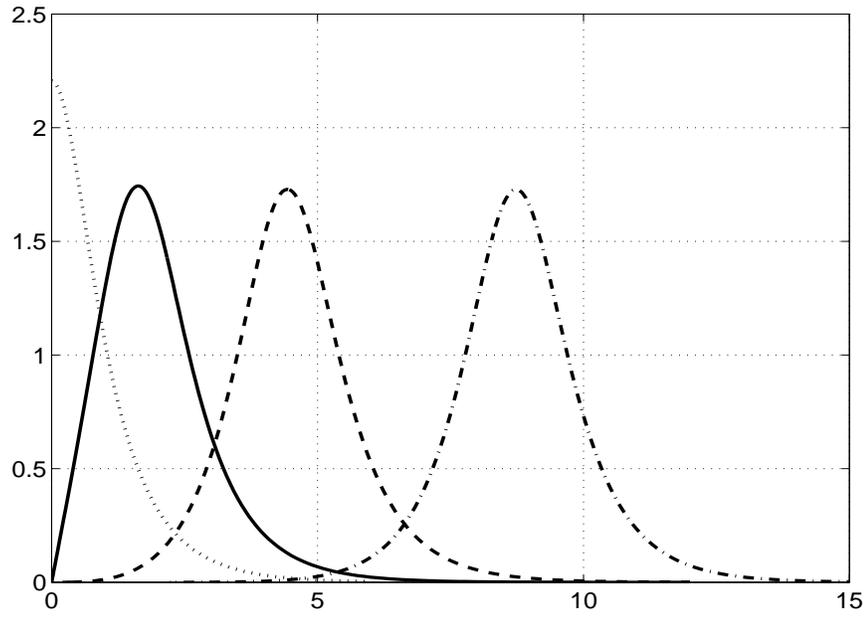}
} } \end{center} 
\end{figure}

{\it Cubic.}
In the case of a cubic nonlinearity $g(u)=\gamma|u|^2u$ with
$\gamma>0$, we may always scale so that $\gamma=1$ and $\omega=1$, by
replacing $w(r)$ by $\sqrt{\omega/\gamma}\, \tilde w(\sqrt\omega
r)$. 
Figure~\figref{FCubicVarySpin} shows the nodeless ($n=0$)
wave profiles for spin values $m=0$, 1, 3 and 6. 

The form of the nodeless waves varies with $m$ in a way that can be
understood as follows.
It appears that as $m$ becomes large, $w(r)$ remains small
initially on a large interval, and then is approximated near its
maximum at $r=R$ by $\phi_a(r-R)$, where $\phi_a$ is the even positive
localized solution on $(-\infty,\infty)$ of
\be
\phi''-a^2\phi+\phi^3 = 0.
\label{2wAeq}
\ee
with $a=\sqrt{1+(m/R)^2}$. Explicitly,
$\phi_a(x)=\sqrt{2}\,a\sech^2ax$. We will argue in appendix A
that $R\approx\sqrt2 m$, so that $\max
\phi_a(x)=\sqrt3$. This approximation agrees well with the
numerical solution shown in Fig.~\figref{FCubicVarySpin}.

\begin{figure} 
\caption{Nodeless profiles for cubic-quintic 
\label{FCQVarySpin}
with $\omega=0.18$, varying spin $m$}
\begin{center} \leavevmode { \hbox{
        \epsfxsize=\xsz
        \epsfysize=\ysz
        \epsffile{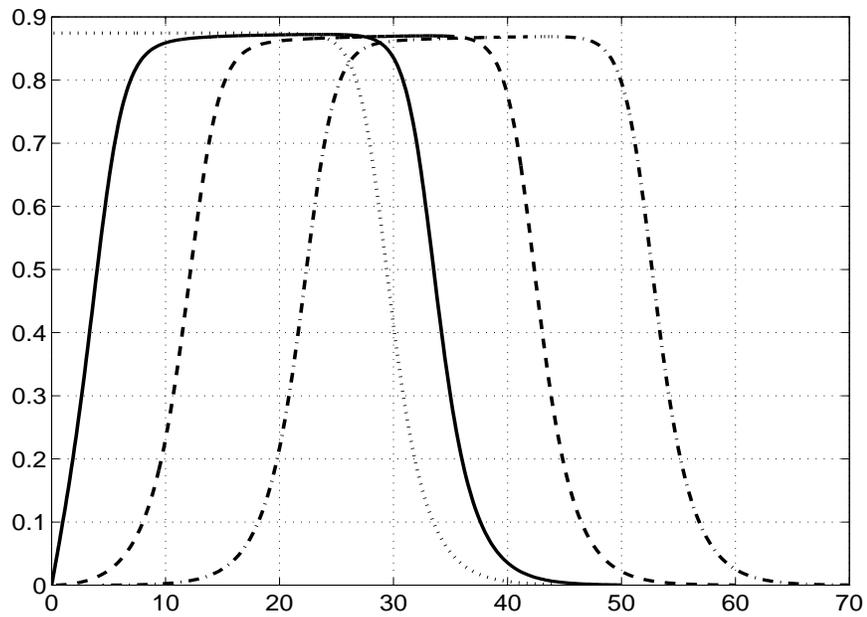}
} } \end{center} 
\end{figure}

{\it Cubic-quintic.}
In the case of a cubic-quintic nonlinearity $g(u)=\gamma|u|^2u
-\delta|u|^4u$ with positive constants $\gamma$ and $\delta$, we can
scale so there is one free parameter $\omega >0$, and
$\gamma=\delta=1$. The condition that $F(w)>0$ for some $w$ is then
equivalent to the requirement that
\be
\omega<\omegas\equiv\frac3{16}=0.1875.
\label{2om*}
\ee
When this holds, $F(w)$
achieves its maximum value at a point $w=\wmax>0$ which depends on
$\omega$. Potentials $F$ with this additional local maximum are
called hilltop potentials in \ccite{IW3} and \ccite{IWW}.  We write
$\Fmax=F(\wmax)$ and note that $\wmax^2=\half+\half\sqrt{1-4\omega}$.
For $\omega=\omegas$ we get $\wmax=\wmaxs\equiv\sqrt3/2$.
Figure~\figref{FCQVarySpin} shows nodeless profiles with $\omega=0.18$
for spin values $m=0$, 1, 2, 3. 

\begin{figure} 
\caption{Nodeless profiles with $m=1$ for cubic-quintic, varying $\omega$}
\label{FCQVaryOmegam1}
\begin{center} \leavevmode { \hbox{
        \epsfxsize=\xsz
        \epsfysize=\ysz
        \epsffile{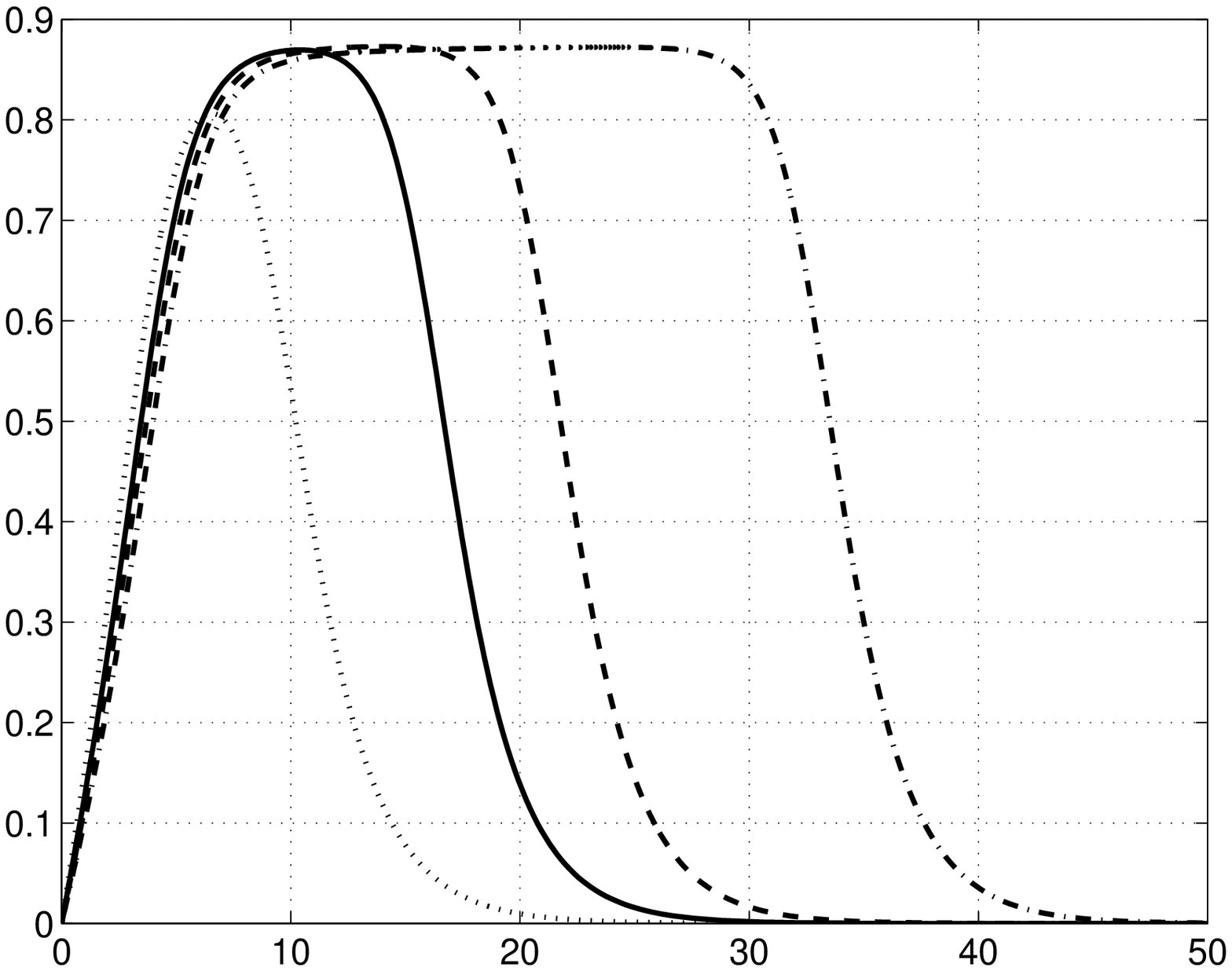}
} } \end{center} 
\end{figure}

\begin{figure} 
\caption{Nodeless profiles with $m=3$ for cubic-quintic, varying $\omega$}
\label{FCQVaryOmegam3}
\begin{center} \leavevmode { \hbox{
        \epsfxsize=\xsz
        \epsfysize=\ysz
        \epsffile{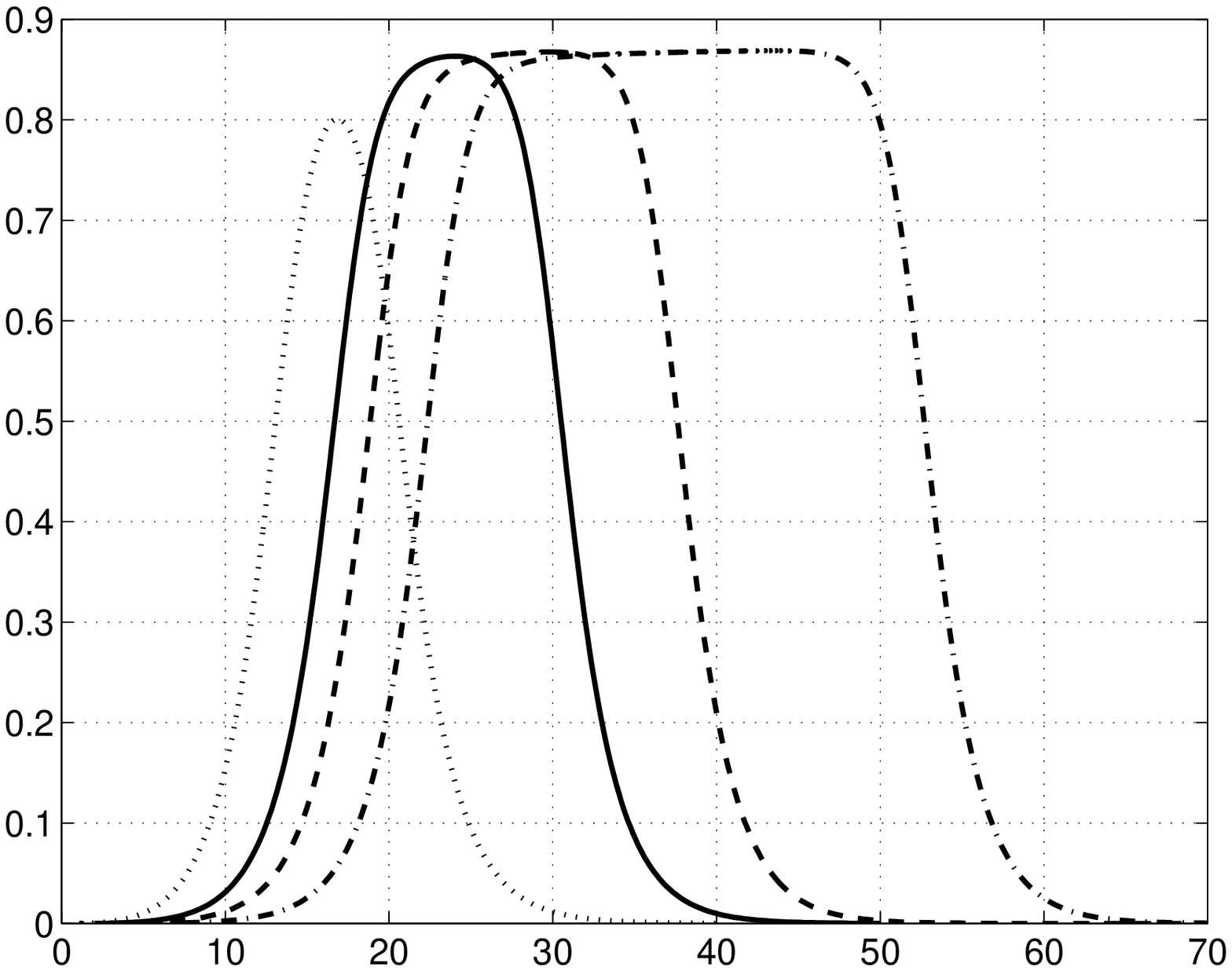}
} } \end{center} 
\end{figure}

We demonstrate the effect of varying $\omega$ in
Figures~\figref{FCQVaryOmegam1} and \figref{FCQVaryOmegam3}, where we plot
nodeless profiles for spins $m=1$ and 3, respectively.  As $\omega$
approaches $\omegas$ from below, the height $\Fmax$ of the rightmost
hilltop in the graph of $F$ decreases to zero, and the ``particle'' at
position $w(r)$ spends more ``time'' near $\wmax$, resulting in flatter
profiles.  This is the phenomenon of ``loitering at the hilltop''
studied in \ccite{IW3} and \ccite{IWW}, also see \cite{Coleman}. 
It is precisely in this ``flattop'' regime that we find
nodeless waves with no unstable eigenvalues for any $m$, 
as discussed in Section \ref{Result}.

The wave profiles develop a structure in this regime that can be
described more precisely.
It is shown in \ccite{IW3} that for any $\omega\in(0,\omegas)$, with an
appropriate amplitude $d$ in (\ref{2asymp}) one obtains a
nonlocalized solution of (\ref{1weq}) that increases monotonically
to $\wmax$ as $r\to\infty$. This yields a vortex solution to
(\ref{1nls}). Note that for $\omega$ near $\omegas$, the profiles in
Figures~\figref{FCQVaryOmegam1} and \figref{FCQVaryOmegam3} resemble a
vortex profile cut off by a kink.  Near the kink location 
at $r=R$ (where $w(r)$ last equals $\half \wmax$, say), the profile is
approximated by $\phis (r-R)$, where $\phis $ is the solution on
$(-\infty,\infty)$ of
\be
\phi'' +  g(\phi)-\omegas w=0
\label{2kink}
\ee
that satisfies 
\be \label{2kinkbc}
\phis (r)\to\wmaxs \hbox{ as $r\to-\infty$}, 
\quad \phis (r)\to0 \hbox{ as $r\to\infty$}, \quad
\phis (0)=\half\wmaxs.
\ee
We note that a one-dimensional kink solution of \eq{1nls} 
of the form $u(x,y,t)=e^{i\omegas t}\phis(x)$ exists
for any nonlinearity with $h\in C^1$ 
that has the properties (i) and (ii) stated in the introduction.

Fig.~\figref{FCQVarySpin} suggests also that as the spin $m$ increases, the
wave structure contains a flipped inner kink that also moves out away 
from the origin. 
We formally analyze the asymptotic behavior of the kink locations as 
$\omega\to\omegas$ in appendix A.
\section{Linearization and reduction to ordinary differential equations}
\reseteqnos

In this section  we linearize \eq{1nls} about
a given exponentially localized solution of the form \eq{1sw} and prove a
number of fundamental results that concern the associated spectrum 
and eigenvalues. In particular, we prove in Theorem~\ref{Plargej} that the
search for unstable eigenvalues reduces to the study of a finite
number of eigenvalue problems for $2\times2$ systems of second-order
ordinary differential equations. 

We consider a solution $u(x,t)$ to \eq{1nls} that results from a perturbation
to the initial condition $u_0(x,0)=v_0(x)$ for a standing wave solution
$u_0(x,t)=e^{i\omega t}v_0(x)$ where $v_0(x)=e^{im\theta}w(r)$  
with $w(r)\to0$ exponentially as $r\to\infty$.
We write 
\be
u(x,t)=e^{i\omega t}\left( {v_0(x)+v(x,t)} \right)
\label{(4)}
\ee
and linearize the nonlinear terms in the evolution equation for the
perturbation $v$ around amplitude $v=0$. We compute that
\be
g(v_0+v)-g(v_0)= \alpha (w)v+
\beta (w)e^{2im\theta }\bar v+O(|v|^2)
\label{(8)}
\ee
where
\be \label{3abdef}
\alpha(s) = \frac12\left[ {g'(s)+{1 \over s}g(s)} \right], \quad
\beta(s) = \frac12\left[ {g'(s)-{1 \over s}g(s)} \right]
\ee
for $s\ne 0$, with $\alpha(0)=\beta(0)=0$. 
Here $g'$ refers to the ordinary derivative of the real-valued
restriction of $g$ to real arguments. 
Note that because $g$ is not holomorphic, its linearization is
not complex-linear but only real-linear.
Also note that 
$g'(s) = O(s^2)$, so $\alpha(s)=O(s^2)$ and
$\beta(s)=O(s^2)$ as $s\to 0$.

Thus the linearized evolution equation for the perturbation $v$ may be
written
\be
-i \,\partial_t v=\Delta v-\sigma ^2 v+\alpha (w)v+
\beta (w)e^{2im\theta }\bar v
\label{(9)}
\ee
where $\sigma^2=\omega>0$.
We next represent the linearized evolution equation as a pair of real
equations for the components of
\be
{\Phi}\equiv \left( {\matrix{{\Phi_1}\cr
{\Phi_2}\cr
}} \right)\equiv \left( {\matrix{{{\rm Re\,} v}\cr
{{\rm Im\,} v}\cr
}} \right)\in {\R}^2,
\label{(10)}
\ee
so that $v = \Phi_1 +i \Phi_2$.
Setting
\be
J \equiv \pmatrix{0& -1\cr 1& 0},
\quad R \equiv \pmatrix{1&0\cr0&-1},
\label{(12)}
\ee
we rewrite equation \eq{(9)} as
\be
{\partial_t}{\Phi}=A{\Phi}\equiv J\left( {\Delta -\sigma ^2+\alpha
(w(r))+\beta (w(r))e^{2m\theta J}R} \right){\Phi}.
\label{Adef}
\ee

\subsection{Spectrum and eigenvalues} 
As a first step toward understanding the stability properties of the standing
waves, we investigate the spectrum of $A$, regarded as an unbounded operator on
the space $L^2(\R^2,\R^2)$ of square-integrable functions 
with values in $\R^2$. The domain of $A$
is the Sobolev space $H^2(\R^2,\R^2)$ of functions into $\R^2$ whose derivatives
up to order 2 are square-integrable. This space is complexified to study the
spectrum of $A$, which we denote by $\spec(A)$. The
spectrum of $A$ consists of two parts, the discrete spectrum $\specdisc(A)$,
consisting of isolated eigenvalues of finite multiplicity, and the essential
spectrum $\specess(A)$, which consists of everything else and is 
determined by the ``operator at infinity'' defined by 
\[
A_0 = J(\Delta-\sigma^2).
\]
{\lemma \label{spec}
The spectrum of $A$ is the union of the discrete and essential spectrum.
The latter is purely imaginary and is given by 
\[
\specess(A) = \spec(A_0) = \{it\mid t\in\R \mbox{\ and\ } |t|\ge\sigma^2\}.
\]
}

\proof  The spectrum of $A_0$ is easily determined using the Fourier
transform. It consists of the set of all eigenvalues of the matrix
$J(-\xi^2-\sigma^2)$ as $\xi$ varies over $\R$. This set is equal to 
$\{it\mid t\in\R \mbox{\ and\ } |t|\ge\sigma^2\}$.

The operator $A-A_0$ is a multiplier with coefficients
that are continuous and bounded and decay exponentially as $r\to\infty$. 
Consequently, it is easy to show (using the convenient 
compactness criterion of \ccite{Pego85}, for example)
that $A$ is a relatively compact perturbation of $A_0$, meaning that for
$\lambda\notin\spec(A_0)$ the operator $(\lambda-A_0)^{-1}(A-A_0)$ is compact.
From Weyl's theorem, it follows that $A$ and $A_0$ have the same essential
spectrum.
\eproof

{\remark \label{const-stab}
\rm We briefly digress to consider the linearization of \eq{1nls}
about the spatially constant solution $u_0(t)=ae^{i\omega t}$.
One may assume $a>0$ without loss of generality, and 
one has $\omega=g(a)/a$ and $\alpha(a)-\omega=\beta(a)=a^2h'(a^2)$.
Linearization yields the constant-coefficient operator 
$ A_a = J(\Delta+\beta(a)+\beta(a)R), $
whose spectrum is the set $\{\pm\lambda(\xi)\mid \xi\in\R^2\}$ 
where $\lambda(\xi)=i\sqrt{(|\xi|^2-\beta(a))^2-\beta(a)^2}$,
which is purely imaginary if and only if $\beta(a)\le0$. 
Hence the solution $ae^{i\omega t}$ is spectrally stable if and only
if $h'(|a|^2)\le0$.
}

\vskip .2 in
By Lemma \ref{spec}, our primary concern becomes the discrete spectrum
of $A$ in \eq{Adef}.
We will search for eigenfunctions ${\Psi}$ of
$A$ taking values in $\C^2$.  
The operator $A$ has real coefficients and has the special form $A=JL$ where $L$
is self-adjoint. Consequently eigenvalues occur with Hamiltonian symmetry: If
$\lambda$ is an eigenvalue with eigenfunction $\Psi$, then $\bar\lambda$ and
$-\lambda$ are eigenvalues. This is because $\bar {\Psi}$ is
also an eigenfunction of $A$ with eigenvalue $\bar \lambda$, and $J\Psi$ is 
an eigenfunction of the adjoint $A^*=-LJ$ with eigenvalue $-\lambda$.
Note that if $\lambda$ is complex with nonzero imaginary part, then the real and
imaginary parts of $e^{\lambda t} {\Psi}$
are {\it real} solutions to the differential equation \eq{Adef} that are 
linearly independent. 

The equation $A{\Psi}=\lambda {\Psi}$ is equivalent to 
\be
(\lambda J-\sigma^2+\Delta+\alpha){\Psi}
 +\beta e^{2m\theta J}R{\Psi} =0.
\label{(14)}
\ee
Suppose $\Psi$ is a square-integrable solution of \eq{(14)}. 
We represent ${\Psi}(r,\theta)\in {\C}^2$ via eigenvectors of
$J$:
\be
{\Psi}= \psi_+\pmatrix{1/2\cr -i/2} + \psi_-\pmatrix{1/2\cr i/2},
\label{(15)}
\ee
where $\psi_+(r,\theta)$ and $\psi_-(r,\theta)$ are complex-valued.
Since
\be
J\Psi = i \psi_+\pmatrix{1/2\cr -i/2} -i \psi_-\pmatrix{1/2\cr i/2}
,\quad 
R\Psi = \psi_-\pmatrix{1/2\cr -i/2} + \psi_+\pmatrix{1/2\cr i/2} ,
\label{(15b)}
\ee
taking the components of \ref{(14)} gives
\be \ba{rcl}
(i\lambda-\sigma^2+\Delta+\alpha)\psi_+ +\beta e^{2im\theta}\psi_-&=&0 ,
\\
(-i\lambda-\sigma^2+\Delta+\alpha)\psi_- +\beta e^{-2im\theta}\psi_+&=&0.
\ea
\label{(16)} 
\ee
Since $\psi_+$ and $\psi_-$ are square-integrable and $\alpha(w(r))$,
$\beta(w(r))$ are bounded, it follows that $\psi_+$, $\psi_-$ lie in the Sobolev
space $H^2(\R^2)$. In particular, they are H\"older-continuous functions.

We next expand $\psi_+$ and $\psi_-$ in Fourier series with respect to
$\theta$:
\be
\psi_\pm (r,\theta )=\sum\limits_{n=-\infty }^\infty  
{\,\,e^{in\theta }y_\pm^{(n)} (r)},
\label{(19)}
\ee
with
\be
y_\pm^{(n)} (r)={{1 \over {2\pi }}}\int_0^{2\pi } 
{\,\,\psi_\pm (r,\theta )e^{-in\theta }\,\,d\theta }.
\label{(20)}
\ee
This is an orthogonal expansion and each term lies in $H^2(\R^2)$.
It will prove notationally convenient to shift the summation index 
and write instead
\be
\psi_\pm (r,\theta )=\sum\limits_{j=-\infty }^\infty  
{\,\,e^{i(j\pm m)\theta}y_\pm^{(j\pm m)} (r)}.
\label{(17)}
\ee
The Fourier coefficients then satisfy
\be
\ba{rcl}
(i\lambda-\sigma^2+\Delta_r - r^{-2}(j+m)^2+\alpha)y_+^{(j+m)} 
+\beta y_-^{(j-m)}&=&0,
\cr
(-i\lambda-\sigma^2+\Delta_r - r^{-2}(j-m)^2+\alpha)y_-^{(j-m)} 
+\beta y_+^{(j+m)}&=&0,
\ea
\label{ysys}
\ee
and satisfy the boundary conditions
\be
\lim_{r\to0^+}y_\pm^{(j\pm m)} (r) \ \ \mbox{exists, \quad and}\quad
\lim_{r\to\infty}y_\pm^{(j\pm m)} (r) =0.
\label{ysysbc}
\ee
Here $\Delta_r\equiv{\textstyle {\D^2\over\D r^2}+{1\over r}{\D\over\D r}}$ 
is the radial Laplacian.

For each integer value of $j$, equations (\ref{ysys}) are a system of
second-order ordinary differential equations for the two functions
$y_+^{(j+m)}(r)$ and $y_-^{(j-m)}(r)$.  There is no
coupling between these systems for different values of $j$.
Summarizing, we have proved the ``only if'' part of the following:

\begin{proposition} \label{Psomej}
{ A complex number $\lambda\notin\specess(A)$ is an eigenvalue of $A$
in $L^2$ if and only if for some integer $j$, 
equations \eq{ysys} have a nontrivial solution satisfying \eq{ysysbc}. }
\end{proposition}

To prove the ``if'' part, we claim that if there are nontrivial solutions
$y_\pm^{(j\pm m,\lambda)}$ to equations \eq{ysys}--\eq{ysysbc} 
for some $\lambda$ and some $j$, then
\be 
{\Psi}^{(j,\lambda)} =\left( {\matrix{1/2\cr -i/2\cr}} \right)
e^{i(j+m)\theta}y_+^{(j+m,\lambda)}(r)+
\left( {\matrix{1/2\cr i/2\cr}} \right)
e^{i(j-m)\theta}y_-^{(j-m,\lambda)}(r)
\label{evecdef}
\ee
is an eigenvector of $A$ with eigenvalue $\lambda$. 
Clearly $\Psi={\Psi}^{(j,\lambda)}$ satisfies \eq{(14)} except perhaps 
at the origin. We must show that $\Psi\in H^2(\R^2,\R^2)$. 
It follows from \eq{2asymp} and
the theory of asymptotic behavior of ODEs (see ~\ccite{Coppel} and
Subsection~\ref{Asym} below) that solutions of \eq{ysys} that decay to
zero as $r\to\infty$ must decay at an exponential rate, together with
their derivatives. Therefore $\Psi\in L^2(\R^2,\R^2)$. 
Furthermore, solutions that approach a limit as $r\to0$ must have 
bounded derivatives.  It follows that $\Psi$ satisfies 
\eq{(14)} in the sense of distributions, and from this it follows that
$\Psi\in H^2(\R^2,\R^2)$. This proves the Proposition.
\eproof

Corresponding to the complex eigenvector \eq{evecdef} 
of the operator $A$ are the two real-valued solutions
\be
{\Phi}_{(a)}^{(j,\lambda)}(t) \equiv 
{\rm Re\,}( e^{\lambda t} {\Psi}^{(j,\lambda)}), \quad
{\Phi}_{(b)}^{(j,\lambda)}(t) \equiv 
{\rm Im\,}( e^{\lambda t} {\Psi}^{(j,\lambda)})
\label{(23)}
\ee
of the vector evolution equations (\ref{Adef}), from which we
recover two complex-valued solutions of the original linearized equation
(\ref{(9)}) via
\be
v^{(a)}(t) = \Phi_{(a)1}^{(j,\lambda)}(t) + i \Phi_{(a)2}^{(j,\lambda)}(t),\quad 
v^{(b)}(t) = \Phi_{(b)1}^{(j,\lambda)}(t) + i \Phi_{(b)2}^{(j,\lambda)}(t). 
\label{(24)}
\ee
Explicitly, we have:
\be
\ba{rcl}
v^{(a)}(t)&=&\sfrac12\left( e^{\lambda t} e^{i(j+m)\theta }y_+
+ \overline{ e^{\lambda t} e^{i(j-m)\theta}y_-}\right),
\\   
v^{(b)}(t) &=&\sfrac{1}{2i}\left( e^{\lambda t}e^{i(j+m)\theta}y_+
- \overline{ e^{\lambda t} e^{i(j-m)\theta}y_-}\right).
\ea
\label{(25)}
\ee
Since any real linear combination of $v_{(a)}$ and $v_{(b)}$ is also a
solution, the general solution to the linearized equation \eq{(9)}
corresponding to eigenvalue $\lambda$ and index $j$ is 
\be
v(x,t) = \left( c e^{\lambda t} e^{i(j+m)\theta }y_+^{(j+m,\lambda)}(r)
  + \overline{c e^{\lambda t} e^{i(j-m)\theta } y_-^{(j-m,\lambda)}(r)
}\,\,\right)
\label{(26)}
\ee
where $c$ is any complex constant.
If we were to take the time-zero value of this function $v(x,0)$ as a
perturbation to the initial conditions $v_0(x)$ for \eq{1nls}, we would
expect the nonlinear time evolution of $v$ to 
be well-approximated by (\ref{(26)}) for small time.

We note that $j$ measures the departure of the perturbation $v$ from the angular dependence of $u_0$.  We will refer to $j$ as the ``twist'' index.

\subsection{Bounds on index for unstable eigenvalues}\label{Largej}

We have shown eigenvalues of the operator $A$ correspond to eigenvalues of
\eq{ysys}-\eq{ysysbc} for some value of the index $j$. 
To locate eigenvalues of the operator $A$,  we use numerical
techniques to locate eigenvalues of \eq{ysys}--\eq{ysysbc}. 
This procedure at first
appears intractable because there is a countably infinite set of equations to
analyze.  
Our main result in this section, however, establishes that any eigenvalue with
nonzero real part occurs in a system (\ref{ysys}) with twist $j$ in a finite
set determined by the wave profile $\ws$ alone.

\begin{theorem}\label{Plargej}
Suppose that $\Re\lambda\ne0$, 
where $\lambda$ is an eigenvalue of $A$ corresponding to a
solution of \eq{ysys}--\eq{ysysbc} for some integer $j$. Then 
\[
|j| \le \jmax :=|m|+ \sqrt{C_2}, \quad \mbox{where}\quad
C_2  = \max_{r\ge 0} 
r^2\left( {\alpha (\ws(r))+\left| {\beta (\ws(r))} \right|} \right).
\]
\end{theorem}

\proof
Let $\mu(s) =\alpha(s) + |\beta(s)|$.  Then $\mu(0)=0$ and $\mu(s) = \sigma^2
+\max\{ f'(s), f(s)/s \}$ for $s\ne 0$. Because $f$ is continuously
differentiable, $\mu$ is continuous on $\R$, and we note that $\mu(r)$ takes on
positive values because $f(w(r))$ does (otherwise the primitive $F(w(r))$ would
never get positive). Because $\mu(s) = O(s)$ as $s\to 0$ and $\ws(r)$ is
exponentially decaying at $\infty$, the maximum $C_2$ of $r^2 \mu(\ws(r))$ is
achieved, and we have $0<C_2<\infty$.

We note that with $y_\pm=y_\pm^{(j\pm m)}$, (\ref{ysys}) can be written
\be
L_j {\bf y} = \lambda K{\bf y},
\label{KLeq}
\ee
where
\be
{\bf y}= \pmatrix{y_+\cr y_-},\quad
K = \pmatrix{i& 0\cr 0& -i},
\ee
and
\be
L_j= \pmatrix{\sigma^2-\Delta_r + r^{-2}(j+m)^2-\alpha & -\beta \cr
-\beta & \sigma^2-\Delta_r + r^{-2}(j-m)^2-\alpha}.
\ee
Define inner products in $L^2(\R^+;r\,dr)$ and $L^2(\R^+,\R^2; r\,dr)$ by 
\[
\ipbig{y,z} =  \int_0^\infty y(r)\overline{z(r)}\,r\,dr, \quad
\ipbig{\bfy |\bfz} = \ipbig{y_+,z_+}+\ipbig{y_-,z_-}.
\]

We claim that if 
$C_*:=\min\{(j-m)^2,(j+m)^2\}> C_2$ and \eq{KLeq} has a nontrivial solution,
then  $\ip{L_j\bfy |\bfy}>0$. To see this, recall that for a solution of
\eq{ysys}--\eq{ysysbc}, $y_\pm$ and its derivative are bounded as
$r\to0$ and decay exponentially as $r\to\infty$. We then have that
$\ipbig{(\sigma^2-\Delta_r)y_\pm , y_\pm}\ge0$, therefore
\begin{eqnarray*}
\ipbig{L_j\bfy|\bfy} &\ge& 
\int_0^\infty \left\{ \left(\frac{j+m}{r}\right)^2 |y_+|^2 
+\left(\frac{j-m}{r}\right)^2|y_-|^2 \right\} r\,dr \\
&& -\ \int_0^\infty \Bigl\{ \alpha\left(|y_+|^2+|y_-|^2\right) + 
\beta \left( y_+\bar{y}_-+\bar{y}_+y_-\right) \Bigr\} r\,dr \\
&\ge & \int_0^\infty (C_*-C_2)\left(|y_+|^2+|y_-|^2\right) r^{-1}\,dr
>0,
\end{eqnarray*}
where we used the facts that ${\bar y}z+y{\bar z}\le |y|^2+|z|^2$ and
$\alpha+|\beta|\le C_2/r^2$.

Now take the inner product of \eq{KLeq} with $\bfy$. 
Since $C_*=(|j|-|m|)^2$,
if $|j|>|m|+\sqrt{C_2}$ we get
\[
0< \ipbig{L_j\bfy|\bfy} = \lambda \ipbig{K\bfy|\bfy}.
\]
Since $K$ is skew-adjoint, $\ipbig{K\bfy|\bfy}$ is purely imaginary,
therefore $\lambda$ must be purely imaginary.
\eproof

\subsection{Bounds on eigenvalues}

We show here that eigenvalues of $A$ are confined to a vertical strip
determined by the wave amplitude profile $\ws$.

\begin{proposition} \label{L.eigbd}
{If $\lambda$ is an eigenvalue of $A$, then
\[
|\Re\lambda|\le \Lmax := 
\max_{r\ge 0} \Bigl(|\alpha(w(r))|+|\beta(w(r))|\Bigr).
\]
}\end{proposition}

\proof Let $\tilde A=A-A_0 = J(\alpha+\beta e^{2m\theta J}R)$. 
Since the $\ell^2$ matrix norms $|J|$, $|R|$ and $|e^{2m\theta J}|$
all equal $1$, we have 
$|\tilde A|\le C_0 := \max_{r\ge 0} (|\alpha(w(r))|+|\beta(w(r))|)$.
Note that $\lambda-A_0$
is the Fourier multiplier $\calF(M)$ defined via 
$(\calF(M)\Psi)\hat{\ }(\xi)=M(\xi)\hat\Psi(\xi)$,
where
\be
M(\xi)= \pmatrix{\lambda &-\sigma^2-|\xi|^2\cr \sigma^2+|\xi|^2&\lambda}.
\ee

Suppose $\Psi$ is an 
eigenfunction of $A$ corresponding to an eigenvalue $\lambda$ with nonzero
real part. Then $0=(\lambda-A_0-\tilde A)\Psi$, and $\lambda-A_0=\calF(M)$ 
is invertible, so that we have
\be \label{eigeqM}
(I-\calF(M^{-1})\tilde{A})\Psi = 0.
\ee
Let $\pdis_\pm = \lambda\pm i(\sigma^2+|\xi|^2)$. Then
$\lambda= \half (\pdis_++\pdis_-)$, 
$\sigma^2+|\xi|^2 = \frac1{2i}(\pdis_+-\pdis_-)$,
so 
\[
M(\xi)= Q\diag\{\pdis_-,\pdis_+\}Q^{-1}, \qquad
Q = \frac1{\sqrt2}\pmatrix{1&1\cr i&-i}.
\]
Note that $Q$ is unitary and 
$M(\xi)^{-1}=Q\diag\{\pdis_-^{-1},\pdis_+^{-1}\}Q^{-1}$,
so the matrix norm 
$|M(\xi)^{-1}|\le \max{|\pdis_\pm^{-1}|} \le 1/|\Re\lambda|$.
Therefore, the operator norm of 
$\calF(M^{-1})\tilde{A}$ on $L^2$ is bounded by $C_0/|\Re\lambda|$.
If $|\Re\lambda|>C_0$, then, there is no nontrivial solution of 
\eq{eigeqM} in $L^2$.
\eproof

\subsection{Symmetries}\label{EvSym}

The value $\lambda=0$ is always an eigenvalue of the operator $A$, 
with multiplicity greater than one, due
to various symmetries of the original nonlinear Schr\"odinger
equation \eq{1nls} (e.g., translation, rotation, phase change). 
These symmetries and associated generalized
eigenfunctions are described in detail in Appendix~\ref{ZeroModes}.

The eigenvalue equations \eq{ysys} also have  simple symmetries
connected to the Hamiltonian symmetry of $A$. From \eq{ysys} and the
orthogonal expansion in \eq{(17)} it is easy to check the following.

\begin{lemma} Suppose $(y_+,y_-)$ form a solution to \eq{ysys} for
some pair $(\lambda,j)$. Then $(y_-,y_+)$ form a solution for
$(-\lambda,-j)$ and $(\bar y_-,\bar y_+)$ form a solution for
$(\bar\lambda,-j)$. If $\lambda$ is a real eigenvalue of $A$
corresponding to index $j\ne0$, then $\lambda$ is (at least) a double
eigenvalue.
\end{lemma}

\section{Detecting eigenvalues with Evans functions}
\reseteqnos

In this section we prove that eigenvalues of the
operator $A$ of index $j$ correspond to zeros of a certain analytic
function $E_j(\lambda)$, which we call an Evans function for the 
system \eq{ysys}.  
This characterization of eigenvalues as zeros of an analytic function
allows us to use the argument principle to count the total number of eigenvalues
inside various contours in the complex plane.

We shall provide three equivalent expressions for this Evans function,
in terms of a Wronskian,  an adjoint system, and an associated exterior system. 
The last expression is used to establish the global analyticity 
of $E_j(\lambda)$ and is advantageous for reasons of numerical stability.

We develop the theory in this section only for the case of spinning
waves for which $m\ne0$. The spinless case $m=0$ is different in
several ways, primarily due to the nonzero value of the wave profile 
$w(r)$ at $r=0$. For brevity's sake we
omit any theoretical treatment of the spinless case, though in section 5
we report results of numerical computations performed for this case.

\subsection{Asymptotic behavior of eigenfunctions.}\label{Asym}

Here we study the asymptotic behavior of solutions to
\eq{ysys} as $r\to 0^+$ and as $r\to\infty$.  
We fix the integer $m\ne0$, the base wave radial profile $\ws$, 
the index $j$, and a complex number
$\lambda$ in the cut plane $\Ccut=\C\setminus\{i\tau:
|\tau|\ge\sigma^2\}$.

Because $\ws(r)\to 0$ exponentially as $r\to\infty$, for large values of $r$
the coefficients $\alpha(\ws(r))$ and $\beta(\ws(r))$ in
\eq{ysys} vanish exponentially quickly as $r\to\infty$.  We thus
expect that solutions to \eq{ysys} will behave for large $r$ like
solutions of the decoupled equations
\be 
(i\lambda-\sigma^2+\Delta_r - r^{-2}(j+m)^2)y_+ =0,
\label{ysysp} 
\ee 
\be
(-i\lambda-\sigma^2+\Delta_r - r^{-2}(j-m)^2)y_- =0. 
\label{ysysm}
\ee
Solutions of \eq{ysysp} are linear combinations of the modified Bessel
functions $I_{|j+m|}(k_+ r)$ and $K_{|j+m|}(k_+ r)$, and solutions of
\eq{ysysm} are linear combinations of the modified Bessel
functions $I_{|j-m|}(k_- r)$ and $K_{|j-m|}(k_- r)$,
where
\be \label{41kpm}
k_+= \sqrt {\sigma ^2-i\lambda}, \quad 
k_-= \sqrt {\sigma ^2+i\lambda},
\ee
with the principal branch of the square root taken.  
For brevity we write
\be \label{IKdef}
\ba{c}
I_+(r)=I_{|j+m|}(k_+ r),\quad I_-(r)=I_{|j-m|}(k_- r),\cr 
K_+(r)=K_{|j+m|}(k_+ r),\quad K_-(r)=K_{|j-m|}(k_- r).
\ea
\ee
Since $\lambda$ is
in the cut plane, the complex numbers $k_\pm$ have positive real parts,
and thus $|I_{\pm}(r)|\to\infty$ and 
$|K_{\pm}(r)|\to 0$ as $r\to\infty$.

Similarly, because $\ws(r)\to 0$ as $r\to 0^+$ when $m\ge 1$, we expect
that solutions to \eq{ysys} will also behave in this case like solutions
of the decoupled equations \eq{ysysp}, \eq{ysysm} for $r$ near zero.
Note that $|I_{\pm}(r)|$ is bounded and 
$|K_{\pm}(r)|\to \infty$ as $r\to0$.

To make these observations precise, we reformulate \eq{ysys} as a
first-order system. We set
\be
c_+(r)= k_+^2+ \left(\frac{j+m}{r}\right)^2, \quad 
c_-(r)= k_-^2+ \left(\frac{j-m}{r}\right)^2.
\ee
Then \eq{ysys} is equivalent to the system
\be
\yf' = \Bf(r,j,\lambda)\yf
\label{41yf}
\ee
with $\Bf(r,j,\lambda)=\halo{\Bf}(r,j,\lambda)+\tBf(r)$, where,
writing $y_+=y_+^\jpm$, $y_-=y_-^\jmm$, we have
\be
\label{41Bdef}
\yf=\pmatrix{y_+ \cr {y_+}' \cr y_- \cr {y_-}'} ,\quad
\halo{\Bf} = \pmatrix{0&1&0&0\cr c_+& -1/r &0&0\cr 0&0&0&1\cr
0&0&c_-&-1/r}, \quad 
\tBf = \pmatrix{0&0&0&0\cr -\alpha&0&-\beta&0\cr 0&0&0&0\cr
-\beta&0&-\alpha&0} .
\ee

Let $\hyf_i$, $i=1,\ldots,4$, denote the linearly independent
solutions 
of the asymptotic system
\be
\yf' = \halo{\Bf}(r,j,\lambda)\yf
\label{yasymsys}
\ee
that form the columns of the fundamental matrix
\be
\hY = \pmatrix{
I_+(r) &0&K_+(r)&0 \cr
I_+'(r)&0&K_+'(r)&0 \cr
0& I_-(r) &0&K_-(r) \cr
0& I_-'(r)&0&K_-'(r) } .
\label{yhalodef}
\ee

{\lemma \label{4yasym}
Fix $\lambda$ in the cut plane $\Ccut=\C\setminus\{i\tau: |\tau|\ge\sigma^2\}$,
and let $m>0$.
Then there exist solutions $\yf_i^{(0)}$ and $\yf_i^{(\infty)}$ to the
coupled system \eq{41yf} such that
\bee
\label{41y0asy}
\yf_i^{(0)}(r) &\sim& \hyf_i(r) \ \ \hbox{as}\ \ r\to 0^+, \\
\label{41yiasy}
\yf_i^{(\infty)}(r) &\sim& \hyf_i(r) \ \ \hbox{as}\ \ r\to \infty,
\eee
for $i=1,2,3,4$.
}

The proof of Lemma \ref{4yasym} is in Appendix \ref{Pf4.1}.

\subsection{Evans function test for eigenvalues}\label{Wronski}

We fix the spin $m\ge1$, base wave $\ws$, index $j$, and a complex number
$\lambda$ in the cut plane $\Ccut$.  
To determine whether $\lambda$ is an eigenvalue corresponding to
index $j$, we wish to determine whether \eq{ysys} has a nontrivial
solution satisfying \eq{ysysbc}. This is equivalent to determining
whether \eq{41yf} has a bounded solution that vanishes as $r\to\infty$.

Let $\yf_i^{(0)}$ and $\yf_i^{(\infty)}$, $i=1,\ldots,4$, be solutions
of \eq{41yf} with the properties specified in Lemma~\ref{4yasym}.
From Lemma \ref{4yasym} we know that the subspace
$S_0$ of solutions $\yf$ to \eq{41yf} that are bounded as $r\to 0^+$
is spanned by $\yf_1^{(0)}$ and $\yf_2^{(0)}$. 
Also we know that the subspace $S_\infty$ of solutions $\yf$ to \eq{41yf} 
that approach zero as $r\to\infty$ is spanned by
$\yf_3^{(\infty)}$ and $\yf_4^{(\infty)}$.

The question of whether there are nontrivial solutions of \eq{ysys}
satisfying \eq{ysysbc} for the value of $\lambda$ under consideration
is thus reduced to the question of whether the subspaces $S_0$ and
$S_{\infty}$ have nontrivial intersection. To answer this question,
we may compute the Wronskian of the four solutions 
$\yf_1^{(0)}$, $\yf_2^{(0)}$,
$\yf_3^{(\infty)}$ and $\yf_4^{(\infty)}$. Since there
is a nontrivial intersection of $S_0$ and $S_{\infty}$ if and only if
these four vectors are linearly dependent, the Wronskian
of these four vectors vanishes (identically in $r$)
if and only if the number $\lambda$  is an eigenvalue corresponding
to index $j$.  

We note that, since $\trace(\Bf)=-2/r$, by Abel's formula the quantity 
\be \label{42Wdef}
E_j(\lambda):= -r^2
{\det\left(\yf_1^{(0)},\yf_2^{(0)},\yf_3^{(\infty)},\yf_4^{(\infty)}\right)}
\ee 
is independent of $r$. This normalized Wronskian is called the 
{\it Evans function} \cite{AGJ} associated with the system \eq{41yf}.
Thus we have the following.

\begin{proposition} \label{4Wzero}
A complex number $\lambda$ 
in the cut plane $\Ccut$
is an eigenvalue of $A$ if and only if $E_j(\lambda)=0$ for some 
integer $j$.
\end{proposition}

We conclude this subsection by describing some symmetries of this
Evans function.

{\begin{proposition} \label{4Esym}
For all integers $j$ and complex numbers $\lambda\in\Ccut$,
\begin{itemize}
\item[(i)] $\overline{E_j(\lambda)}=E_j(-\overline{\lambda})$
\item[(ii)] $E_j(\lambda)$ is real if $\lambda$ is purely imaginary.
\item[(iii)] $E_j(\lambda) = E_{-j}(-\lambda)$
\end{itemize}
\end{proposition}
}

\proof
(i) Since taking the complex conjugate of \eq{41yf} yields a solution of 
the same equation with $\lambda$ replaced by $-\overline{\lambda}$,
we find from \eq{42Wdef} that 
$\overline{E_j({\lambda})} = E_j(-\overline{\lambda})$. 

(ii) follows from (i), since $\lambda=-\overline{\lambda}$ if
$\lambda$ is purely imaginary.

(iii)  We make use of the symmetry of equations \eq{ysys} as
described in subsection~\ref{EvSym}. If $\yf$ is a solution of
\eq{41yf}, then exchanging the first two components with the last two
yields a solution of $\yf' = \Bf(r,-j,-\lambda)\yf$. Consequently,
$E_{-j}(-\lambda)$ is obtained from $E_j(\lambda)$ in \eq{42Wdef} by
exchanging the first two rows with the last two.  This leaves the
determinant invariant, and then (iii) follows.
\eproof

\subsection{The adjoint system}

To obtain alternative characterizations of the Evans function,
we consider next the adjoint system associated with \eq{41yf},
\be
\zf' = -\zf \Bf(r,j,\lambda).
\label{4zf}
\ee
Here $\zf(r)$ is considered to be a row vector. Any solutions of
\eq{4zf} and \eq{41yf} have the property that the product
$\zf(r)\cdot\yf(r)$ is independent of $r$.  The asymptotic adjoint
system
\be
\zf' = -\zf \halo{\Bf}(r,j,\lambda)
\label{4zf0}
\ee
has the fundamental set of solutions $\halobz_k$, $k=1,\ldots,4$ given by
the rows of the matrix
\be
\hZ(r) = r\pmatrix{
-K'_+(r)& K_+(r) &0&0\cr
0&0&-K'_-(r)& K_-(r) \cr
I'_+(r)& -I_+(r) &0&0\cr
0&0&I'_-(r)& -I_-(r) }
\label{4haloZ}
\ee
This matrix is the inverse of $\hY(r)$, i.e., we have
$\hZ\hY=I$.

{\lemma\label{4zasym} The adjoint system \eq{4zf} has solutions $\zf^\rt_k$,
$k=1,\ldots,4$ that satisfy 
\[
\zf^\rt_k(r)\sim\halobz_k(r) \quad\mbox{ as
$r\to\infty$.}
\]
}
This lemma is proved in the same manner as Lemma \ref{4yasym}, so
we omit the details.

Let  $\zf_k^{(\infty)}$, $k=1,\ldots,4$, be solutions 
with the properties specified in Lemma~\ref{4zasym}.
Let $Z^\rt(r)$ denote the matrix whose rows are $\zf^\rt_k(r)$,
and let $Y^\rt(r)$ be the matrix with columns
$\yf^\rt_k(r)$. Because of the limiting behavior as $r\to\infty$
expressed in Lemmas~\ref{4yasym} and \ref{4zasym} and the fact that
$\hZ\hY=I$, it follows directly that
\be
Z^\rt(r) Y^\rt (r)= I
\label{4ZYrt}
\ee
for all $r>0$; that is, $\zf^\rt_k(r)\cdot\yf^\rt_l(r)=\delta_{kl}$.

The matrix $Y^\rt$ is a fundamental matrix for the system \eq{41yf}.
Therefore the solutions $\yf^\lt_i$ of Lemma~\ref{4yasym} admit
representations of the form
\be
\yf^\lt_i = \sum_{k=1}^4 \yf^\rt_k\alpha_{ki}
\label{4yrelate}
\ee
with constant coefficients that are given by
\[
\alpha_{ki}= \zf^\rt_k\cdot\yf^\lt_i.
\]
The value $\lambda$ is an eigenvalue of \evpj if and only if
the span of $\yf^\lt_1$ and $\yf^\lt_2$ nontrivially intersects the span
of $\yf^\rt_3$ and $\yf^\rt_4$. This is equivalent to the statement that
\be
\det\pmatrix{\alpha_{11}&\alpha_{12}\cr\alpha_{21}&\alpha_{22}}
=\det\pmatrix{ \zzyy{1}{1}&\zzyy{1}{2}\cr \zzyy{2}{1}& \zzyy{2}{2}} =0.
\label{4Edef1}
\ee
It turns out that this determinant simply yields an 
alternative characterization of the Evans function.

{\begin{proposition} \label{4EW}
For all integers $j$ and complex numbers $\lambda\in\Ccut$,
\[
E_j(\lambda)=
\det\pmatrix{ \zzyy{1}{1}&\zzyy{1}{2}\cr \zzyy{2}{1}& \zzyy{2}{2}}.
\]
\end{proposition}
}

\proof
\[
\ba{rcl}
E_j(\lambda)&=&-r^2
{\det\left(\yf_1^{(0)},\yf_2^{(0)},\yf_3^{(\infty)},\yf_4^{(\infty)}\right)}\cr
&=&-r^2\sum\limits_{i,k=1}^4\alpha_{i1}\alpha_{k2}
{\det\left(\yf_i^{(\infty)},\yf_k^{(\infty)},
\yf_3^{(\infty)},\yf_4^{(\infty)}\right)}\cr
&=&-r^2\left(\alpha_{11}\alpha_{22}-\alpha_{21}\alpha_{12}\right)
{\det\left(\yf_1^{(\infty)},\yf_2^{(\infty)},
\yf_3^{(\infty)},\yf_4^{(\infty)}\right)}\cr
&=&-r^2\det\left(\bmY^{(\infty)}\right)(\alphadet) \cr
&=&-r^2\det\left(\bmY^{(0)}\right)(\alphadet) \\[10pt]
&=& \det\pmatrix{ \zzyy{1}{1}&\zzyy{1}{2}\cr \zzyy{2}{1}& \zzyy{2}{2}}.
\ea
\]
Here we have used the fact that the $r$-independent quantities 
$r^2\det\left(\bmY^{(\infty)}\right)$ and $r^2\det\left(\bmY^{(0)}\right)=-1$
are equal by virtue of \eq{41yiasy}.
\eproof

\subsection{Exterior systems and analyticity}

According to our results so far, for each fixed $\lambda$ in
$\Ccut$, $E_j(\lambda)$ depends upon a choice of certain solutions
$\yf^\lt_i$ and $\yf^\rt_i$ or $\zf^\rt_i$ having the properties
described in Lemmas~\ref{4yasym} and \ref{4zasym}.  
In this section, we 
establish
that the value of $E_j(\lambda)$ is {\it independent} of this
choice, and that $E_j$ is an analytic function in the 
cut plane $\Ccut$.

The method of proof of
Lemmas~\ref{4yasym} and \ref{4zasym}
could be used to establish that the required solutions can be chosen to
depend analytically on $\lambda$, locally in the set $\Ccut$ 
in the complex plane.
However, this approach does not 
guarantee that there is a single
globally analytic choice, leaving the difficulty that a given
set of choices may yield an Evans function that is not analytic 
in all of $\Ccut$.  

This difficulty is obviated with an algebraic device described by
Alexander and Sachs \ccite{AS} (used by one of the present authors 
for numerical calculations mentioned in that paper) and recently
employed by Afendikov and Bridges \ccite{AB}. 
A similar device (called the compound matrix shooting method) has been 
used in numerical computations for boundary-value problems 
related to hydrodynamic stability theory \ccite{NR}. 

One considers {\it exterior products} of solutions of the system
\eq{41yf} and the associated adjoint system \eq{4zf}.
In the present context it suffices to consider exterior products of two
solutions at a time. 

In general, suppose $\yf$ and $\tyf$ are vectors in $\C^n$ that,
with respect to the standard basis vectors $\bme_j$, $j=1,\ldots,n$,
have the expansions
\be
\yf = \sum_{j=1}^n y_j\bme_j,\quad
\tyf = \sum_{j=1}^n \tilde{y}_j\bme_j.
\label{4yexp}
\ee
Let $N={n\choose2}=n(n-1)/2$ and let $\ywedgety$ denote the vector
in $\C^N$ with components
\be
(\ywedgety)_{j\wedge k} = y_j\tilde{y}_k-y_k \tilde{y}_j
= \det\pmatrix{ y_j&\tilde{y}_j\cr y_k&\tilde{y}_k}
\label{4ywedge}
\ee
for $j<k$. The vector $\ywedgety=-\tyf\wedge\yf$ is called the
exterior product of $\yf$ and $\tyf$.
For convenience, when $n=4$ we order the components according
to the identification
\[ (1\wedge2,\ 1\wedge3,\ 1\wedge4,\ 2\wedge3,\ 2\wedge4,\ 3\wedge4)=
(1,2,3,4,5,6).\]

Suppose that $\yf$ and $\tyf$ are both solutions of the linear system
\be
\yf' = \Bfour \yf
\label{4By}
\ee
where the matrix-valued $\Bfour$ has columns $\bmb_j$, $j=1,\ldots,n$.
Then since $\Bfour\yf=\sum y_j\bmb_j$, one computes that
$\ysix:=\ywedgety$ satisfies the linear system
\be
\ysix' = (\Bsix) \ysix
\label{4hatB}
\ee
where matrix $\Bsix$ has columns given by
$\BM{\hat{b}}_{j\wedge k} = \bmb_j\wedge\bme_k+\bme_j\wedge\bmb_k$.
The system in \eq{4hatB} is called the {\it exterior system} associated
with \eq{4By}.  Similarly, for two solutions $\zf$ and $\tzf$
of the adjoint system
\[
\zf' = -\zf\Bfour
\]
the exterior product $\zsix:=\zwedgetz$ is a solution to the associated exterior
system
\be
\zsix' = -\zsix (\Bsix) .
\ee
It is a straightforward matter to compute that
\be
(\zwedgetz)\cdot(\ywedgety) = \det\pmatrix{
\zf\cdot\yf&\zf\cdot\tyf \cr \tzf\cdot\yf&\tzf\cdot\tyf}.
\label{4zywedge}
\ee

As a consequence of \eq{4zywedge} and Proposition~\ref{4EW}, 
the Evans function $\Evans_j$ may be expressed as
\be
\Evans_j(\lambda) = 
(\zrwzr)\cdot(\ylwyl).
\label{4Edef2a}
\ee
This expression has advantages for both theoretical and
numerical purposes, which stem from the fact that 
the exterior products $\zrwzr$ and $\ylwyl$ 
have a separate characterization  
that shows they are independent of any particular
choice of the individual solutions involved.

{\begin{proposition} \label{4Eana}
Fix an integer $m\ne0$, the base wave radial profile $\ws$, 
the index $j$, and a complex number $\lambda\in \Ccut$.
The exterior system associated with \eq{41yf}
has a unique solution $\ysix^\lt(r,\lambda)$with the asymptotic behavior
\[
\ysix^\lt \sim   
\hylwyl \quad\mbox{as $r\to0^+$.}
\]
Similarly, the exterior system associated with the adjoint system \eq{4zf}
has a unique solution $\zsix^\rt(r,\lambda)$ with the asymptotic behavior
\[
\zsix^\rt \sim   
\hzrwzr \quad\mbox{as $r\to\infty$.}\]
Furthermore, the solutions $\ysix^\lt(r,\lambda)$, $\zsix^\rt(r,\lambda)$
are analytic functions of $\lambda$ globally in the cut plane $\Ccut$. 
\end{proposition}}

\begin{corollary} \label{4Ezy}
For any choice of solutions $\yf^\lt_k$, $\yf^\rt_k$, $\zf^\rt_k$ 
having the properties indicated in 
Lemmas~\ref{4yasym} and \ref{4zasym} we have
\[
\yf^\lt_1\wedge\yf^\lt_2= \ysix^\lt 
,\quad
\zf^\rt_1\wedge\zf^\rt_2 = \zsix^\rt .
\]
The Evans function $E_j$ is given by
\be
E_j(\lambda) = \zsix^\rt\cdot\ysix^\lt,
\label{4Edef2}
\ee
and $E_j$ is analytic globally in the cut plane $\Ccut$.
\end{corollary}

The proof of Proposition \ref{4Eana} is in Appendix \ref{PfAna}.
The uniqueness property of the solution $\ysix^\lt$
with the given asymptotic normalization 
stems from the fact that it is a solution of the 
exterior system associated with \eq{41yf}
having {\it maximal decay rate} as $r\to0$.
Similarly, $\zsix^\rt$ is a solution of the adjoint
exterior system having maximal decay rate as $r\to\infty$.

Corollary~\ref{4Ezy} follows from Proposition~\ref{4Eana} via the 
observation that $\yf^\lt_1\wedge\yf^\lt_2$ and 
$\zf^\rt_1\wedge\zf^\rt_2$ are solutions of the exterior systems 
associated with \eq{41yf} and \eq{4zf}, respectively,
with the asymptotic behaviors given in Proposition~\ref{4Eana}.
Thus both $\yf^\lt_1\wedge\yf^\lt_2$ and 
$\zf^\rt_1\wedge\zf^\rt_2$ are globally analytic functions of
$\lambda$ in $\Ccut$, and it follows from \eq{4Edef2a} that $E_j$
is given by \eq{4Edef2} and is also globally analytic in $\Ccut$.


\section{Numerical computation and results} \label{Result}
\reseteqnos

\subsection{Evaluation of 
Evans functions}\label{NumWron}

To evaluate the Evans function $E_j(\lambda)$ in \eq{42Wdef}
for given $j$ and $\lambda$, we use a Runge-Kutta ODE package (rksuite) to 
compute numerical approximations to the solutions
$\yf_1^{(0)}(r)$ and $\yf_2^{(0)}(r)$ of \eq{41yf} that are bounded at
the origin.  To do this, we solve
two initial value problems for the system \eq{41yf} 
on an interval $r_0\le r\le r_\infty$, where $r_0$ and $r_\infty$
are chosen so that $w(r)$ is small (usually $<10^{-4}$) on
$(0,r_0)$ and $(r_\infty,\infty)$. Then $\alpha(w(r))$ and $\beta(w(r))$ 
are close to their asymptotic values on
the intervals $(0,r_0)$ and $(r_\infty,\infty)$.  
We make the approximations $\yf_i^{(0)}(r_0)\approx \hyf_i(r_0)$ 
($i=1,2$) to determine the initial conditions.
We then evaluate the Wronskian of the computed approximations to
$\yf_1^{(0)}(r_\infty)$ and $\yf_2^{(0)}(r_\infty)$ with the
explicitly known approximations $\hyf_3(r_\infty)$ and
$\hyf_4(r_\infty)$ to the respective values $\yf_3^{(\infty)}(r_\infty)$
and $\yf_4^{(\infty)}(r_\infty)$ of the solutions to
the system \eq{41yf} that vanish at infinity.

Our actual numerical implementation of this scheme makes use of a
change of variables and scaling to reduce the sometimes rapid variation of
solutions and improve the quality of the computation.  
We represent a solution $\yf(r)$ of the system \eq{41yf} as
an $r$-dependent scaled linear combination 
of the basis elements  $\left\{\hyf_i:i=1,\dots,4\right\}$ 
for the asymptotic system \eq{yasymsys}, writing  
\be
\yf(r)=\hY(r)\exp(\bmK r)\bmb(r)
\label{yYa}
\ee
where $\bmK:={\rm diag}\{-k_+,-k_-,k_+,k_-\}$,
Then we can anticipate that the vector $\bmb(r)$ is well-behaved
as $r$ becomes large, and it satisfies the system
\be
\bmb'(r)=\bmD(r)\bmb,
\label{asys}
\ee
whose coefficient matrix
$\bmD(r)=-\bmK+e^{-K r}{\hY}{}^{-1}(r)\tBf(r)\hY(r)e^{K r}$ remains 
bounded as $r\to\infty$. 

The Evans functions $E_j(\lambda)$ are also approximated in a similar but 
simpler way based on the expression in \eq{4Edef2}, 
by computing an approximation to $e^{-kr}\ysix^\lt(r)$
for $r_0\le r\le r_\infty$ and using the known asymptotic approximation 
for $\zsix^\rt(r_\infty)$. The scaling factor $k$ is given by $k=k_++k_-$.
Numerical convergence studies indicate that we usually attain 8 to 10
digit accuracy with this approach.
This method of evaluating $E_j(\lambda)$ 
is very stable, since $\ysix^\lt$ is a solution of 
the exterior system with maximal asymptotic growth rate.
It also avoids a potential difficulty
one can encounter with the Wronskian approach --- 
maintaining the linear independence of 
approximations to $\yf^\lt_1$ and $\yf^\lt_2$ is difficult
if the growth rates for these solutions differ greatly and if the
interval of computation is large.

\subsection{Counting eigenvalues with the argument principle}\label{Roundup}

Let us first recall what is known about the possible 
location of unstable eigenvalues.
According to Proposition \ref{Psomej} we may locate eigenvalues of $A$ by
locating zeros of the Evans functions $E_j(\lambda)$.  For a given standing
wave $u_0=e^{i\omega t}e^{im\theta}w(r)$, Proposition \ref{L.eigbd}
guarantees that all such zeros lie in the strip where 
\be
|\Re\lambda|\le \Lmax
:= \max_{r\ge 0} (|\alpha(w(r))|+|\beta(w(r))|).
\ee
Theorem \ref{Plargej} guarantees that all unstable eigenvalues $\lambda$
will be zeros of $E_j$ with 
\be
|j|\le j_{\rm max}:=|m|+
\max_{r\ge 0}\sqrt{r^2({\alpha (\ws(r))+
\left| {\beta (\ws(r))} \right|})}\ .
\ee
Furthermore, it follows from Proposition \ref{4Esym} 
that $E_{-j}(\lambda)=0$ if and only if
$E_j(\bar\lambda)=0$, so that zeros of $E_{-j}$ in the strip
$0<\Re\lambda\le \Lmax$ are in one-to-one correspondence with zeros of $E_j$
in that strip.

\begin{figure} 
\caption{Schematic of contour used for counting eigenvalues}
\label{ContourH}
\begin{center} \leavevmode { \hbox{
        \epsfxsize=\xszz
        \epsfysize=\ysz
        \epsffile{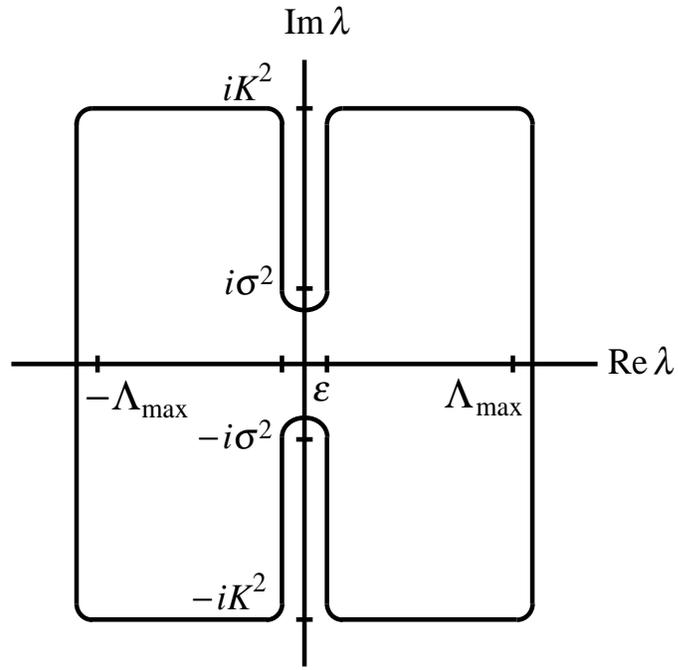}
} } \end{center} 
\end{figure}

To search for zeros of the Evans functions in the strip
$|\Re\lambda|\le \Lmax$, we apply the argument principle to count the
number of zeros of the analytic functions $E_j$ inside a contour that
encloses a ``large" part of the strip. We do not have a rigorous bound
on $|\Im\lambda|$ for unstable eigenvalues, but in practice we find
that $|E_j(\lambda)|$ grows for large $|\lambda|$, so we take a large
contour determined so that an asymptotic trend appears established.
Figure~\figref{ContourH} shows a schematic of the contour $\gamma$
employed in our computations, lying in the cut plane $\Ccut$. The
contour is designed to enclose the region where $|\Re\lambda|\le\Lmax$
and $|\Im\lambda|\le K$, except for an $\epsilon$-neighborhood of the
essential spectrum $\{it\mid |t|\ge\sigma^2\}$. The parameter
$\epsilon$ is taken small ($\approx 10^{-6}$), and $K$ large.  

Using the computed approximation to $E_j(\lambda)$ for given $j$ and
$\lambda$ described above, we compute the winding number of the image
$E_j(\gamma)$ about 0. This is the change in the argument of the
complex number $E_j(\lambda)$ as $\lambda$ traverses the curve
$\gamma$, divided by $2\pi$. We use an adaptive stepping procedure to
traverse $\gamma$ in steps that result in a small relative change in
$E_j(\lambda)$, and accumulate the change in ${\rm arg} E_j(\lambda)$.
By the argument principle, this yields the number of zeros of $E_j$
enclosed by $\gamma$. 

Because of the symmetry $E_j(-\bar\lambda)=\overline{E_j(\lambda)}$ from
Proposition \ref{4Esym}(iii), in practice we need only to compute the
change in argument of $E_j(\lambda)$ along the portion $\gamma_r$ of the
contour $\gamma$ in the right-half plane $\Re\lambda\ge 0$.  Doubling
yields the total change in argument along $\gamma$.

By the same symmetry, the number of zeros enclosed by $\gamma$ in the
right half-plane is the same as the number in the left half-plane. 
So, from the winding number above, we need to subtract the number of purely
imaginary zeros and divide by two to determine the number of {\em
unstable} eigenvalues of $A$ corresponding to index $j$ enclosed by
$\gamma$. To find the purely imaginary zeros we exploit the fact
established in Proposition~\ref{4Esym}
that $E_j(\lambda)$ is real if $\lambda$ is purely imaginary.
We plot the real function
$E_j(it)$ for $t\in[-\sigma^2+\epsilon,\sigma^2-\epsilon]$, refining and
checking the multiplicity of zeros with log-log plots as necessary.

The computation of the winding number can be considerably slowed by the
presence of zeros of $E_j(\lambda)$ very near the contour --- the
adaptive stepping procedure takes small steps to track the image curve
near such a zero.  The reason we do not take our contour to be a
simple rectangle with one side along $\Re\lambda=\epsilon$ is to
avoid passing near numerous zeros on the imaginary axis between
$-i\sigma^2$ and $i\sigma^2$.

Nevertheless, it sometimes happens that the values of $E_j(\lambda)$ become
small as $\lambda$ traverses the part of the contour $\gamma_r$
in the right half-plane near the essential spectrum on the
imaginary axis. This behavior
suggests that the contour passes near a zero of $E_j$. To verify that
these zeros do not lie in the right half-plane outside $\gamma$, we use
low-degree least-square polynomial fits to extrapolate an approximation to
$E_j(\lambda)$ near $\gamma_r$, and a root finder to locate the zeros.
In all cases the zeros appear to lie in the left half-plane.
In view of the symmetry $\overline{E_j(\lambda)}=E_j(-\overline{\lambda})$,
we interpret these findings to indicate 
not that $E_j$ has zeros in the cut plane $\Ccut$ outside $\gamma$,
but rather that the analytic continuation
of $E_j(\lambda)$ across the imaginary axis has zeros near the axis.
Such points are known as {\it resonance poles}, see \cite{PW92}.

\subsection{Spinless ground-state waves}\label{StabTh}

The stability of ground state (nodeless) standing-wave solutions 
to \eq{1nls} with
nonlinearities that include those we consider here is analyzed in
\cite{GSS2} and \cite{W86}.
These works establish that a family of ground state standing waves ($m=n=0$),
parametrized by standing wave frequency $\omega$, is 
orbitally stable (stable modulo spatial translations and phase shifts 
under $H^1$ perturbations of initial data) if 
$dN/d\omega>0$
and unstable (to radial perturbations) if
$dN/d\omega<0$,
where $N=\|u_0\|^2$.
For general nonlinearities, 
the stability of the ground state in the marginal case
$dN/d\omega = 0$
is open, but for pure-power nonlinearities
$g(u)=\gamma |u|^{p-1}u$ (for which $dN/d\omega = 0$
is equivalent in $d$ spatial dimensions to 
$p=1+4/d$) the ground state is unstable in the sense
that there exist arbitrarily small $L^2$ perturbations of initial data for
which the corresponding solution of (NLS) blows up in $H^1$ norm in finite
time. (See \ccite{W83}.)

The construction and computation of the Evans functions for these
spinless waves differs from the description we have given above, 
since the wave amplitude $w(r)$ approaches a nonzero value as $r$
approaches zero. Nevertheless, through diagonalization of the coupling
matrix in \eq{ysys}, explicit (formal) asymptotic solutions may be written
and Evans functions computed. For brevity we omit the details,
since our primary concern is the spectral stability of standing waves
with spin.  

We performed computations for spinless waves for the cubic and 
cubic-quintic nonlinearities, both to validate the numerical code and
to study the mechanism of transition to instability as the quantity
$dN/d\omega$ changes sign.

The cubic nonlinear Schr\"odinger equation in two spatial dimensions
is a marginal case from the point of view of theory ($dN/d\omega=0$).
Recall that by scaling we may without loss of generality
take $g(u)=|u|^2u$ and consider only the fixed standing wave
frequency $\omega=1$.
The bounds from Theorem~\ref{Plargej} and Proposition~\ref{L.eigbd} yield that 
$|j|\le\jmax=1.85$ and $|\Re\lambda|\le14.6$ for unstable eigenvalues.
In fact we find no eigenvalues with positive real part,
indicating that the mechanism for instability in this case is not
exponential linear instability.
The eigenvalue $\lambda=0$ has high multiplicity, however --- it
is found to be a zero of $E_j$ of order 4 for $j=0$ and of order 2 for $j=1$.
(This remains true for the cubic for all spin/node-number combinations
considered.) This order is consistent with the enumeration of zero
modes in Appendix B. No other discrete eigenvalues are found. 

To study transition to instability, we perturb the cubic nonlinearity 
and analyze the one-parameter family $g(u)=|u|^2u-\delta |u|^4u$.  
Making use of the scaling transformation
$w(r)=\sqrt{\omega}\tilde{w}(\sqrt{\omega}\,r)$, we may without loss of
generality study eigenvalues for the single standing-wave frequency 
$\omega=1$. For the wave to exist, 
$\delta$ must satisfy $\delta<{3\over 16}$.
Numerical computation of $N=2\pi \int_0^\infty  {w(r)^2r\,dr}$ 
for the wave profiles indicates that, by the stability criterion 
for ground states, $u_0$ is stable ($dN/d\omega>0$)
for $0<\delta <\textstyle{3\over 16}$ and is unstable ($dN/d\omega<0$)
for $\delta<0$.

Our numerical search for eigenvalues shows that the transition to
instability of the ground state as $\delta$ decreases through zero is
characterized by the collision of two purely imaginary conjugate eigenvalues
in the twist-0 ($j=0$) subspace.  We observe the transition by plotting
$\Evans_0(i\tau)$ for $\tau\in (-\sigma^2,\sigma^2)$ for various values of
$\delta$.  For $0<\delta<\textstyle{3\over 16}$, log-log plots 
indicate that the function $\Evans_0(i\tau)$
has a zero of multiplicity two at $\tau=0$ and two other conjugate simple
zeros.  As $\delta\to 0^+$, the simple zeros approach $\tau=0$, and at
$\delta=0$ there is a zero of multiplicity four at $\tau=0$.
For $\delta<0$, two real zeros of $\Evans_0(\lambda)$ emerge
with opposite sign on the real axis. This creates a positive eigenvalue
that renders the standing wave exponentially unstable.

\subsection{Waves with nodes} 
In all our computations for ``first excited states,''
waves whose profiles have a single node 
where $w(r)=0$ for some $r>0$, we find many unstable eigenvalues.
This holds whether the waves have spin or not.
For example, for the cubic with $m=0$, $n=1$, we find 12 
unstable eigenvalues, distributed as 
$(2,1,1,1,1,1,0\ldots)$, which we abbreviate as
$(2,1(5\times),0\ldots)$, meaning
two in index $j=0$ and one each in index $j$ for $|j|=1,\ldots,5$.
For $m=1$, we find 24 unstable eigenvalues, 
distributed as $(2(4\times),1(5\times),0\ldots)$.
For $m=2$, we find 34, distributed as $(2(6\times),1(6\times),0\ldots)$.
The existence of eigenfunctions with twist
$j=0$ leads us to expect that the higher-node-number states
have exponential instabilities to perturbations of initial data of the
form $e^{im\theta}v(r)$.
For the cubic-quintic $g(u)=|u|^2u-|u|^4u$ we only mention a few
cases with $n=1$ since now different values of $\omega$ 
are in principle different: 
with $m=0$, $\omega=0.05$ we find 10 unstable 
eigenvalues;
with $m=1$, $\omega=0.16$ we find 20; 
and with $m=2$, $\omega=0.165$ we find 24.
It proved difficult to perform accurate computations for profiles
with more than one node ($n>1$), but we expect all such waves to be unstable.

\subsection{Spinning waves with no nodes}

For the cubic nonlinearity (scaled so $\omega=1$) our numerical
results indicate that spinning standing waves with no nodes ($n=0$)
are unstable. With spin $m=1$, there are 6 
unstable eigenvalues\footnote{$1.17\pm1.52i$, $1.97\pm1.69i$, $1.81\pm1.88i$}, 
one for each twist index with $|j|=1,2,3$. And
there are 8 eigenvalues in the ``gap'' $\{it\mid
-\sigma^2<t<\sigma^2\}$, distributed in $j$ as $(4,2,1,0,1,0\ldots)$. For
$m=2$, there are 6 unstable, one for each $j=1,\dots,6$, with 11 gap
eigenvalues distributed as $(4,2,1,1,1,0,0,1,1,0\ldots)$.

Next we consider the cubic-quintic nonlinearity
$g(u)=\gamma|u|^2u-\delta|u|^4u$, where the
constants $\gamma$ and $\delta$ are both positive.  
As mentioned in section \ref{tax}, the scaling 
$w(r)=\sqrt{\gamma\over\delta}\, \tilde w({\gamma\over{\sqrt\delta}} r)$
allows us to suppose without loss of generality that $\gamma=\delta=1$.
The allowable range for $\omega$, for which $F(s)>0$ for some $s>0$, is
$0<\omega<\omegas={3\over 16}$. 

{\it Spectrally stable waves.}
For the nodeless waves with spins $m=1,2,3,4,5$, we have
discovered ranges of standing-wave frequency $\omega$ for which all the
Evans functions $\Evans_j$, $j=0,1,\dots,\jmax$, appear to have no zeros with
positive real part.  Our numerical computations indicate that, for a given
value of spin $m$, there is a transition frequency 
$\mum$ such that $A$ possesses
unstable eigenvalues for $0<\omega <\mum$ but $A$ has no unstable
eigenvalues for $\mum<\omega<\omegas$.  
In Table~\ref{T.transit} we list the transition frequency $\mum$, 
the bound on instability index 
from Theorem~\ref{Plargej}
and the computed distribution of all purely imaginary gap eigenvalues. 
The figures in parentheses indicate the number of times a count is
repeated, so that ``$8(3\times),7(6\times)$'' in the last row of the
table means that there are 8 gap eigenvalues for each $j=0,1,2$,
and 7 eigenvalues for each $j=3,4,5,6,7,8$. Accordingly, the total number
of gap eigenvalues for $m=1,2,3,4,5$ respectively is 
22, 44, 78, 154, 268. 
The bound computed from Lemma \ref{L.eigbd} on the real part of any unstable
eigenvalue satisfies $\Lmax<1$ in all cases.
The value $\lambda=0$ is always found to be a double
zero of $E_j$ for $j=0$ and $1$, which corresponds to the four
zero modes enumerated in Appendix B.

\begin{table}\label{T.transit}
\begin{center}
\caption{Eigenvalue distribution and bounds at the instability transition
for the cubic-quintic nonlinearity.
Waves with spin $m$ are spectrally stable for $\mum\le \omega<\omegas$. 
Unstable eigenvalues emerge at $\lamcr$, 
and can exist only for $|j|\le \jmax$ by Theorem \ref{Plargej}.}
\medskip
\begin{tabular}{|c|c|c|c|c|l|}
\hline $m$ &$\mum$&$\lamcr$ & $\jmax$ & \# of zeros in gap,
for $j=0,1,\ldots$ \\\hline
1 & 0.1487 & 0.0478$i$ & 7.66 & $4,3,3,2,1,0\ldots$ \\
2 & 0.1619 & 0.0271$i$ & 14.2 & $4,4,3(4\times),1(4\times),0\ldots$\\
3 & 0.1700 & 0.0136$i$ & 23.4 & $4,4,5,5,4,3(4\times),2,1(5\times),0\ldots$\\
4 & 0.1769 & 0.0063$i$ & 38.3 &
$6(5\times),5(4\times),3(7\times),2,1(7\times),0\ldots$\\
5 & 0.1806 & 0.0033$i$ & 57.6 &
$8(3\times),7(6\times),6,5(4\times),3(10\times),
2(3\times),1(10\times),0\ldots$ \\
\hline \end{tabular}\end{center}
\end{table}

\begin{figure} 
\caption{$E_2(it)$ {\it vs.} $t$ near instability transition 
\label{TransitMech}
for $m=1$, $\omega=.154$ and $.145$}
\begin{center} \leavevmode { \hbox{
        \epsfxsize=\xsz
        \epsfysize=\ysz
        \epsffile{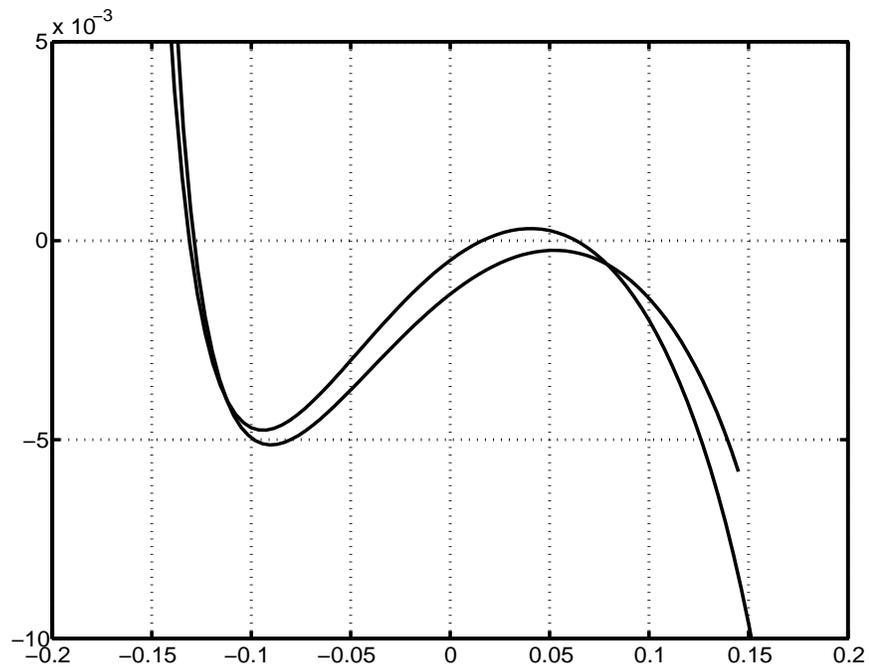}
} } \end{center} 
\end{figure}
\begin{figure} 
\caption{Profiles at stability transition for $m=1,2,3,4,5$}
\label{TransitWaves}
\begin{center} \leavevmode { \hbox{
        \epsfxsize=\xsz
        \epsfysize=\ysz
        \epsffile{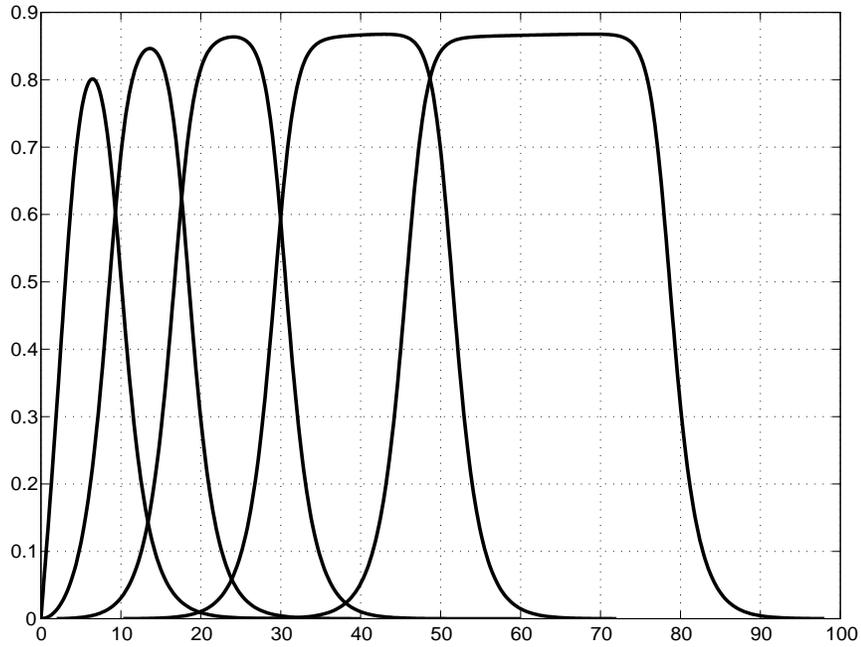}
} } \end{center} 
\end{figure}

For each $m=1,\ldots,5$ the mechanism for transition to instability
as $\omega$ decreases below $\mum$ is the same. A pair of 
purely imaginary gap eigenvalues, always with index $j=2$, 
collides when $\omega=\mum$ at $\lambda=\lamcr$ (as shown in Table 1)
and move away from the imaginary axis in opposite directions, creating
an unstable eigenvalue. Figure~\figref{TransitMech} shows how the
graph of the real-valued function $E_2(it)$ ($-\omega<t<\omega$) 
changes as $\omega$ passes through the critical
frequency for the case of spin $m=1$. 
Figure~\figref{TransitWaves} shows the critical wave profiles for 
$m=1,\ldots,5$.

\begin{figure} 
\caption{Stability transition: 
\label{TransitCurv}
$\log_{10}(\omegas-\omega)$ {\it vs.} $\log_{10}m$ 
for cubic-quintic case}
\begin{center} \leavevmode { \hbox{
        \epsfxsize=\xsz
        \epsfysize=\ysz
        \epsffile{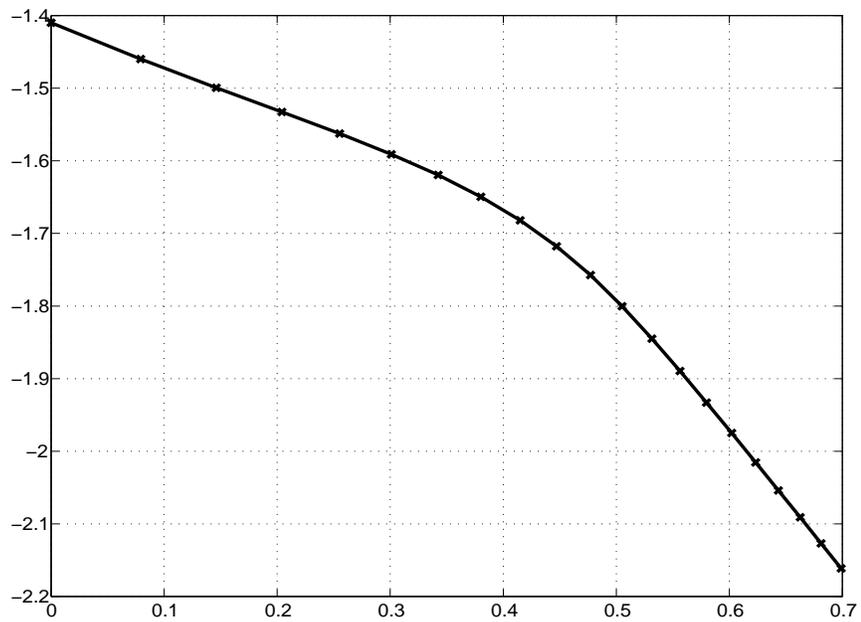}
} } \end{center} 
\end{figure}

Figure~\figref{TransitCurv}
shows a log-log plot of $\mum$ versus $m$. (The
transition frequency is calculated for noninteger values of $m$
by solving the profile equation \eq{1weq} and computing the associated
Evans functions.) From this data it appears that 
\be
\omegas-\mum\sim \frac{0.18}{m^2}
\ee
as $m$ becomes large, suggesting that
stable spinning waves may exist for any spin $m$.

After this paper was submitted, 
the existence of a stability transition frequency $\mum$ for $m=1$ and 2 was
also reported in a paper of Towers {\it et al.}\ \cite{TB2},
who solve numerically the algebraic eigenvalue problem arising from 
discretization of a system of equations equivalent to \eq{ysys}.
These authors also report finding (at the limit of their numerical accuracy)
a weak instability for $j=\pm1$ (see Figure~5 in \cite{TB2}) 
which affects the location of $\mum$ for $m=1$
and $2$ and causes them to find all waves with $m=3$ unstable.
If our present numerical results are correct, 
this weak instability is likely to be a numerical artifact, 
since for $j=1$ the existence of a double eigenvalue at $\lambda=0$
is consistent with the analysis of zero modes in Appendix~\ref{ZeroModes},
and we find no other zeros of $E_1(\lambda)$ inside or near the 
contour $\gamma$ except on the imaginary axis, where $E_j(\lambda)$ is real
and changes of sign clearly locate zeros.

We examined a few cases involving nonlinearities other than
cubic and cubic-quintic to see whether spectrally stable localized standing
waves may exist.
For the focusing-defocusing nonlinearity $g(w)=w^3-w^7$, 
the critical frequency at which $F$ has two
zero-height hilltops is $\omegas\approx 0.2722$. 
Our computations indicate that the wave with 
spin $m=1$ for $\omega=0.24$ is spectrally stable 
($\jmax=9$ and $|\Re\lambda|<0.91$ for unstable eigenvalues in this case).
For the focusing-defocusing nonlinearity $g(w)=w^5-w^7$, 
the critical frequency $\omegas\approx 0.0878$, 
and there is a spectrally stable wave 
with spin $m=1$ for $\omega=0.07$ 
($\jmax=11$ and $|\Re\lambda|<1.27$).

For the saturable nonlinearity $g(w)=w^3/(1+w^2)$, which is always
focusing, we found no spectrally stable spinning waves. 
Spinning solitary waves exist for $0<\omega<1$, growing in
amplitude without bound as $\omega$ increases. With spin $m=1$, 
we find for small $\omega$ that there are three unstable eigenvalues, 
as in the case of the focusing cubic above.
For $\omega\ge0.35$, however, we find only one unstable eigenvalue; 
it has twist index $j=2$. 
And as $\omega$ increases towards 1, the real part of the unstable
eigenvalue becomes small.  E.g., for $\omega=0.5$ the unstable eigenvalue 
is $\lambda=0.174+0.199i$, and for $\omega=0.9$ it is $0.0372+0.0547i$. 
As $\omega$ increases, then, we see that the nonlinearity saturates 
and the wave becomes less unstable. 

\section{Discussion}
\reseteqnos
Our computations support the conjecture that for any
focusing-defocusing nonlinearity, solitary waves with spin
(encapsulated vortices) are spectrally stable for any value of the
spin index $m$ if they have no nodes ($n=0$) and if the standing wave frequency
$\omega$ is sufficiently close to the kink frequency $\omegas$,
which is determined by the property that 
the graphs of $y=g(x)$ and $y=\omegas x$ enclose two regions of
equal area for $0\le x\le \wmaxs$. (Equivalently, the potential
$F=F(s)$ has two zero-height maxima, at $s=0$ and $s=\wmaxs$.)

To begin to understand why these waves might be stable, 
we observe that in the regime when $\omega\approx\omegas$ and $|m|$ is large,
the wave profiles develop a structure that may be accessible to analysis.
This structure is suggested from Figures \ref{FCQVarySpin} and 
\ref{FCQVaryOmegam3}, 
and a formal analysis is given in Appendix A.
In this regime, the wave amplitude $\ws(r)$ is approximately zero
in a large central
core region, then rises quickly in a transition layer where $r\approx\Rin$
to a wide region where $\ws(r)\approx \wmaxs$, then decreases rapidly 
in a transition layer where $r\approx\Rout$ to zero.

Since the wave amplitude remains approximately constant near $\wmaxs$ 
in a wide annular region where $r$ is large and $\theta$ is slowly varying, 
the solution is slowly varying in space. 
Now because the nonlinearity is defocusing for amplitudes near $\wmaxs$
($h'<0$) 
the spatially constant solution $u=\wmaxs e^{i\omegas t}$ is spectrally
stable, by remark \ref{const-stab}.
The zero solution is also spectrally stable, but spatially constant
solutions with amplitude in the focusing range $0<a<a_0$ are
not spectrally stable.

In the zones of rapid transition where $r$ is large and $\theta$
slowly varying, the wave profiles
$\ws(r)$ appear to be well-approximated by a one-dimensional kink 
solution of the nonlinear Schr\"odinger equation \eq{1nls}.
Such a solution has the form $\gamma e^{i\omegas t}\phis(x)$, 
where $\phis$ is as described in section 2 and
where $\gamma$ is a complex constant of unit modulus.
This kink connects the zero solution to the 
solution $\gamma\wmaxs e^{i\omegas t}$.  
It is significant to note that the kink solution itself is
a spectrally stable solution of \eq{1nls}.
Because of the invariance of \eq{1nls}, we may take $\gamma =1$ in
analyzing the stability of $\gamma e^{i\omegas t}\phis$.

\begin{proposition}\label{P.kink}
Assume that $g(u)=h(|u|^2)u$ where $h$ is $C^1$ with $h(0)=0$,
and that properties (i) and (ii) of the introduction hold. 
Then \eq{1nls} admits a one-dimensional kink solution 
$u(x,y,t)= e^{i\omegas t}\phis(x)$, where $\phis$ is the unique solution
of \eq{2kink} and \eq{2kinkbc}, and this solution is spectrally
stable.
\end{proposition}

\proof
The existence and uniqueness of $\phis$ are easy to establish.
To analyze the stability, we perform the linearization of section 3 for 
$u_0(x,y,t) =e^{i\omegas t}\phis(x)$, obtaining 
$\partial_t \Psi = A\Psi$, where now
\be
A= J\left(\Delta-\sigma^2 +\alpha(\phis(x)) + \beta(\phis(x))R\right).
\ee
Taking the Fourier transform in the transverse variable $y$, we obtain 
$\partial_t \hat\Psi = \hat{A}\hat\Psi$ with
$ \hat{A}=-J(L+\ell^2 I)$,
where $\ell$ is the Fourier transform variable and
\be
L=\pmatrix{L_+ & 0 \cr 0 & L_-}
=\pmatrix{-\partial_x^2 -\fs'(\phis) & 0 \cr 
0 & -\partial_x^2 -\fs(\phis)/\phis} .
\ee

Consider the spectrum of $L$.  We know 
$\fs'(0)<0$, $\fs'(\wmaxs)<0$, and $\fs(\wmaxs)=0$, therefore
\[
\lim_{x\to\pm\infty} \fs'(\phis(x)) <0 \quad\hbox{and}\quad 
\lim_{x\to\pm\infty}{\fs(\phis(x))}/{\phis(x)}\le 0.
\]
Hence the essential
spectra of $L_+$ and $L_-$ are contained in $[0,\infty)$.
Each operator $L_+$ and $L_-$ annihilates a function of definite sign:
$$
L_-\phis=0 \quad\quad {\rm and} \quad\quad L_+\phis'=0,
$$
so a standard argument based on the maximum principle implies that $L_+$ and
$L_-$ have no negative eigenvalues.
Thus $L+\ell^2 I$ has no negative eigenvalues, and it is simple to show that
then $J(L+\ell^2)$ has no eigenvalues with nonzero real part.  It follows 
that the spectrum of $A$ is purely imaginary.  Thus the kink
$e^{i\omegas t}\phis(x)$ is spectrally stable, as claimed.
\eproof

\noindent
The focusing-defocusing properties of the
nonlinearity thus permit the existence of a spectrally stable 
one-dimensional kink. 

A simplified description of the wave amplitude
profiles, for spinning solitary waves when $\omega\approx\omegas$ and 
$|m|$ is large, is that they have two pieces:
\begin{itemize}
\item[(i)] 
an outer kink region where $w(r)\approx \phis(r-\Rout)$,  
\item[(ii)] 
an inner kink region where $w(r)\approx \phis(\Rin-r)$.
\end{itemize}
Both the inner and outer kink radii $\Rin$ and $\Rout$ become large,
with $\Rout$ much larger that $\Rin$.
Since both pieces approximate a spectrally stable one-dimensional
kink solution, it becomes plausible that the whole wave can be stable.


\appendix

\section{Asymptotic analysis of wave structure}
\reseteqnos

As indicated in section 2, here we formally analyze the structure of
nodeless localized waves in two limiting regimes, namely (i) when the 
spin $m$ becomes
arbitrarily large, and (ii) when the nonlinearity is of ``hilltop''
type and the  standing wave frequency $\omega$ approaches the critical
value $\omegas$ at which the positive maximum value of the potential
$F$ becomes zero, i.e., becomes equal to the local maximum at $0$.

\subsection{Nodeless wave location for large spin index $m$}\label{BumpLoc}

We observe numerically that the nodeless standing wave profiles $\ws$
for large values of $m$ appear to be copies of a single profile,
shifted further to the right as $m$ increases.  To understand and
characterize this property, we note that $r$ is large where the
wave amplitude is non-negligible.  The term $\ws'/r$ in (\ref{1weq}) is thus
apparently negligible, and we surmise that the quantity ${m^2/r^2}$
is approximately constant where the wave amplitude is non-negligible.

We therefore conjecture that $\ws(r)\approx \phi_A(r-R)$, where $\phi_A$
is a putative positive solution to
\be
\phi''-A \phi +f(\phi)=0
\label{2phieq}
\ee
with maximum at the origin, that vanishes as $r\to \pm \infty$, and
where $A={m^2}/{R^2}$, with $R$ to be determined.  Since $A>0$ and
$f'(0)<0$, the solution $\phi_A$, if it exists,
has exponential decay as $|r|\to\infty$.  Energy considerations show that
such a solution $\phi_A$ will exist provided that the even function
$\FA(s)\equiv F(s)-\half A s^2$ has a positive zero.

To determine $R$ or equivalently $A$, we derive an identity valid for profiles.
Multiplying the equation (\ref{1weq}) by $\ws'$ and rearranging terms, 
we obtain
\be
{1\over r}\ws'^2-{m^2 \over r^3}\ws^2 +
\left[ {1\over 2}\ws'^2 - {m^2 \over 2r^2}\ws^2 +F(w) \right]'=0.
\label{2weqid}
\ee
Because $\ws(r)\sim dr^m$ as $r \to 0^+$, if $m\ge 1$ the quantity in
square brackets vanishes as $r \to 0^+$, and it also vanishes as  $r \to
\infty$.  Integrating (\ref{2weqid}), we find
\be
\int_0^\infty \left( {1\over r}\ws'^2-{m^2 \over r^3}\ws^2 \right)\,dr=0
\label{2wid}
\ee
for $m\ge 1$.

Approximating $r$ in the integrand of (\ref{2wid}) by $R$, we find
\be
A\int_0^\infty \phi_A(r-R)^2\,dr \approx \int_0^\infty \phi_A'(r-R
)^2\,dr.
\ee
Extending the integration to the whole line, we arrive at the
approximation
\be
A\int_{-\infty}^\infty \phi_A(x)^2\,dx \approx \int_{-\infty}^\infty
\phi_A'(x)^2\,dx. \label{2aeq}
\ee

We may compute the values of these integrals in terms of $A$ by
quadrature, as follows.  Assume that $\FA$ has a positive zero, and let
$B$ be the smallest positive zero of the function  $\FA$.  (That is,
$B$ is the solution to $F(B)=\half A B^2$.)  If $\phi(r)$ is a solution
to (\ref{2phieq}), then the energy $E[\phi]=\half \phi'(r)^2+\FA(\phi(r))$
is independent of $r$.  It follows from the facts that $\FA(0)=0$ and
$\phi_A(r)\to 0$ as $r\to\infty$ that $E[\phi_A]=0$ and hence that the
turning point of the solution $\phi_A$ at $r=0$ occurs at amplitude $\phi_A(0)=B$.
Furthermore, we may solve $E=0$ for $\phi'$ to obtain
\be
|\phi_A'|=\sqrt{-2\FA(\phi_A)}.
\ee
By changing variables from $r$ to $p=\phi_A(r)$, we find that
\be
\int_0^\infty \phi_A'(r)^2\,dr=\int_0^B\sqrt{-2\FA(p)}\,dp,
\ee
\be
\int_0^\infty \phi_A(r)^2\,dr=\int_0^B \frac{p^2}{
\sqrt{-2\FA(p)}}\,dp.
\ee
Since $\phi_A$ is an even function we may double these values to obtain
the values of the integrals over the whole line.
The resulting expressions inserted into (\ref{2aeq}) yield an equation that
determines $A$.

As an example, we consider the cubic nonlinearity $F(s)=-s+s^3$.  We have
$\FA(\phi)={1\over 4}\phi^4-\half (1+A)\phi$, hence $B=\sqrt{2(1+A)}$ and
\be
\int_0^\infty \phi_A'(r)^2\,dr=\int_0^{\sqrt{2(1+A)}}
\sqrt{(1+A)p^2-\shalf p^4}\ dp={\textstyle\frac23}(1+A)^{3/2}
 \ee
and
\be
\int_0^\infty \phi_A(r)^2\,dr=\int_0^{\sqrt{2(1+A)}} \frac{p^2}{
\sqrt{(1+A)p^2-\half p^4}}\,dp=2\sqrt{1+A}.
\ee
Therefore, in this example, equation (\ref{2aeq}) yields $A=\half$, and we
predict that the large-$m$ profile will be centered at $R=\sqrt{2}\,m$ and
will have amplitude $B=\sqrt{3}$.  This agrees well with the data shown in 
Fig.~\figref{FCubicVarySpin}.
We remark that in this example the
values of the integrals in (\ref{2aeq}) can also be determined from the
explicit formula $\phi_A(x)=\sqrt{2(1+A)}\sech(\sqrt{1+A}\,x)$.

\subsection{Kink location in the flattop regime}\label{KinkLoc}

Here we study the structure of localized waves for ``hilltop''
potentials such as arise for the cubic-quintic in \eq{1g35}, when
$\omega$ approaches the critical value $\omegas$ from below.
For $\omega$ near $\omegas$, in the cubic-quintic case 
we observe numerically that the nodeless wave
profile $\ws(r)$ remains near $\wmax\approx\wmaxs$ in a large interval,
then decreases rapidly to zero when $r$ is near the last point $R$
where $\ws(R)=\frac12 \wmaxs$.  Near this point the profile 
$\ws(r)$ seems well-approximated by $\phis(r-R)$, where $\phis$ is 
the decreasing kink solution to \eq{2kink} that satisfies \eq{2kinkbc}.

To approximately determine $R$, we return
to the identity (\ref{2weqid}) and integrate from $R-X$ to $R+X$ to obtain
\be
\int_{R-X}^{R+X} \left( {1\over r}\ws'^2-{m^2 \over r^3}\ws^2
\right)\,dr= -\left.\left[ {1\over 2}\ws'^2 - {m^2 \over 2r^2}\ws^2 +F(w) \right]
\right|_{R-X}^{R+X}. 
\label{2wid2}
\ee
We fix $X$ large compared to 1 but independent of $R$.

We first approximate the left side of (\ref{2wid2}).  Because $\ws'(r)$
is nonnegligible only for $r\approx R$, and because $\ws(r)\approx
0$ for $r>R$ and $\ws(r)\approx\wmaxs$ for $r<R$, we have
\be
\int_{R-X}^{R+X} \left( {1\over r}\ws'^2-{m^2 \over r^3}\ws^2 \right)\,dr
\approx
{1\over R}\int_{R-X}^{R+X} \ws'^2\,dr -
m^2(\wmaxs)^2\int_{R-X}^{R}{1\over r^3}\,dr.
\ee
We now approximate $\ws'(r)$ in the integrand by ${\phis}'(r-R)$, and
extend the integration to the whole line.  Setting
\be
T\equiv\int_{-\infty}^\infty {\phis}'(x)^2\,dx=
\int_0^{\wmaxs}\sqrt{-2\Fs(p)}\ dp,
\ee
we find
\be
\int_{R-X}^{R+X} \left( {1\over r}\ws'^2-{m^2 \over r^3}\ws^2 \right)\,dr
\approx
{T\over R}-\half m^2 (\wmaxs)^2
\left(\frac{1}{(R-X)^2}-\frac{1}{R^2}\right).
\label{2LHSapprox}
\ee

We next approximate the right side of (\ref{2wid2}).  Because $\ws(r)\approx
0$ for $r>R$ and $\ws(r)\approx\wmaxs$ for $r<R$, we have
\be
\ba{rcl}
\ds -\left.\left[ {1\over 2}\ws'^2 - {m^2 \over 2r^2}\ws^2 +F(w) \right]
\right|_{R-X}^{R+X} &\approx&
\ds F(\wmaxs)-\half\frac{m^2}{(R-X)^2}(\wmaxs)^2 \\
&=& \ds
\half (\omegas-\omega)(\wmaxs)^2-\half\frac{m^2}{(R-X)^2}(\wmaxs)^2,
\ea
\label{2RHSapprox}
\ee
where we have used the facts that $\Fs(\wmaxs)=0$ and
$F(w)-\Fs(w)=\half(\omegas-\omega)w^2$.

Combining (\ref{2LHSapprox}) and (\ref{2RHSapprox}), we obtain the
quadratic equation
\be
(\omegas-\omega)R^2-\frac{2T}{(\wmaxs)^2}R-m^2=0
\ee
for $R$, which depends only on $\omega$ and
the nonlinearity $\fs$. The solution for R is given by
\be
R\approx \Rout=
\frac{T+\sqrt{T^2+(\wmaxs)^4m^2(\omegas-\omega)}}
{(\wmaxs)^2(\omegas-\omega)}. 
\label{2OuterReq}
\ee

In a similar way, we may approximate the location $R$ of the inner
kink of the nodeless wave profile with $m\ge 2$ for
$\omega\approx\omegas$.  We find 
\be
R\approx \Rin=
\frac{-T+\sqrt{T^2+(\wmaxs)^4m^2(\omegas-\omega)}}
{(\wmaxs)^2(\omegas-\omega)}.  \label{2InnerReq}
\ee
Note that 
\be
\Rout-\Rin=
\frac{2T}{(\wmaxs)^2(\omegas-\omega)},
\ee
independent of $m$. 
For fixed $m\ge 2$ we have 
\be
\Rin\to \frac{(\wmaxs)^2 m^2}{2T} \quad {\rm and} \quad 
\Rout\to\infty
\ee
as $\omega$ approaches $\omegas$ from below.

As an example, we consider the cubic-quintic nonlinearity $f(s)=-\omega
s+s^3-s^5$.  We have $F(s)=-\half\omega s^2 + {1\over 4}s^4-{1\over
6}s^6$, $\omegas=\frac{3}{16}$, $\wmaxs=\sqrt3/2$, and
\be
T=\int_0^{\wmaxs}\sqrt{-2\Fs(p)}\ dp=\int_0^{\sqrt3/2}\sqrt{
\sfrac13 p^6-\sfrac12 p^4+\sfrac3{16}p^2}\ dp=\frac{3\sqrt3}{64}.
\ee
Thus we obtain 
\be
\Rout=
{\frac{{\sqrt{3}} + {\sqrt{3 + 256\,(3/16-\omega)\,{m^2}}}}
   {16\,(3/16-\omega)}},
\ee
\be
\Rin={\frac{{-\sqrt{3}} + {\sqrt{3 + 256\,(3/16-\omega)\,{m^2}}}}
   {16\,(3/16-\omega)}},
\ee
which agrees well with the kink locations for the
numerically computed wave profiles. E.g., for $m=3$, $\omega=0.18$ we get
$\Rout\approx 52$, $\Rin\approx 23$
In the limit $\omega\to\omegas$ note that $\Rin\to 8m^2/\sqrt3$.

\section{Zero modes}\label{ZeroModes}
\reseteqnos

Solitary waves are not stable in the ordinary sense that the
solution to the evolution equation with perturbed initial data differs
little from the unperturbed solution for all time.  This can be seen by
considering, for example, two solitary waves with slightly different
velocities whose initial data are approximately equal but which later
drift apart.  Instead, solitary-wave stability is defined in terms of
the difference between perturbed and unperturbed solutions modulo
symmetries of the evolution equation.  For instance, the two solitary
waves just discussed do remain close modulo spatial translation.

Invariance of an evolution equation under symmetries gives rise to an
eigenvalue-zero eigenspace for the generator of time evolution for the
equation linearized about a solution.  Linear instability due to
generalized eigenfunctions in such a space need not correspond to a
real instability, but rather to ``drift'' that can be cancelled by a
symmetry transformation.  On the other hand, an imaginary eigenvalue
with a generalized eigenfunction (of index larger than 1) that does
not correspond to a symmetry operation would suggest instability.

In this section we make use of symmetries to enumerate the zero modes of
\eq{1nls}, obtaining generalized eigenfunctions of the operator $A$
corresponding to eigenvalue $\lambda=0$.  We identify the same
number of zero modes that we find numerically.

Given a family $u_\tau$ of solutions to \eq{1nls} with smooth dependence on
parameter $\tau$, we can differentiate the identity
\be
-i{\D_t}
u_\tau - \Delta u_\tau = g(u_\tau)
\ee
with respect to $\tau$ to obtain a solution $\tilde u$ to the
corresponding equation linearized about $u_0$:
\be
-i{\D_t} \tilde u - \Delta \tilde u = 
\alpha(|u_0|)\,\tilde u + \beta(|u_0|)\frac{u_0}{\bar u_0}\bar {\tilde u}
\label{zm1}
\ee
where 
\be
\tilde u\equiv \left.{ \D_\tau u_\tau}\right|_{\tau =0}.
\ee

We apply this procedure to various one-parameter families of solutions,
each having $u_0=e^{i\omega \,t}v_0(x)=e^{i\omega \,t}e^{i\,m\theta }w(r)$.
With $\tilde v$ given by $\tilde u = e^{i\omega \,t}\tilde v$,
equation (\ref{zm1}) becomes
\be
\D_t\tilde v = B\tilde v
\label{zm2}
\ee
where $B$ is the real-linear operator given by
\be
Bv\equiv i\left( (\Delta+\alpha(w)-\sigma^2)\,v +
\beta(w)e^{2i\,m\theta}\,\bar{v}\right).
\ee
We note that (\ref{zm2}) is equivalent to equation (\ref{(9)}) 
and that $B$ is the single-component form of the operator $A$.
(Because we are considering the real eigenvalue $\lambda=0$, in this section
we prefer not to work with the two-component form and subsequent
complexification.)

\begin{itemize}
\item Invariance of NLS under phase changes:

We define $u_\tau\equiv e^{i\tau}u_0$.  Because
$g(e^{i\tau}u_0)=e^{i\tau}g(u_0)$ and $u_0$ is a solution of \eq{1nls},
$u_\tau$ is a solution for all real $\tau$.
For this family, $\tilde u = iu_0$ and $\tilde v = iv_0$.  
Since $\tilde v$ is time-independent, we see from (\ref{zm2}) that
$\tilde v$ is an eigenfunction of $B$ with eigenvalue zero.
The corresponding eigenfunction
${\bf \Psi}=(\Re\tilde v,\Im\tilde v)$
of $A$ lies in the $j=0$ subspace, according to (\ref{(15)}), 
(\ref{(20)}), and (\ref{(17)}).

\item Invariance of NLS under rotation:

We set $u_\tau\equiv R_\tau u_0$, where $R_\tau$ is the coordinate
transformation that rotates $u_0$ by angle $\tau$.  Thus 
$u_\tau=e^{i\omega \,t}e^{i\,m(\theta-\tau)}w(r)$, whence 
$\tilde u=-imu_0$ and $\tilde v=-imv_0$.  This eigenfunction $\tilde v$
of $B$ is a real constant multiple of that induced by 
invariance under phase change.

\item Invariance of NLS under time translation:

We set $u_\tau (x,t)\equiv u_0(x,t-\tau)$.  Then $\tilde u=-i\omega u_0$
and $\tilde v=-i\omega v_0$.  This eigenfunction $\tilde v$
of $B$ is a real constant multiple of those induced by 
invariance under phase change and rotations.

\item Invariance of NLS under spatial translation:

Writing $(x,y)$ for the components of the spatial coordinate vector, we
begin by setting $u_\tau (x,y,t)\equiv u_0(x-\tau,y,t)$.  Then 
$\tilde u= \D_x u_0$ and $\tilde v= \D_x v_0$.
Since $\tilde v$ is time-independent, we see from (\ref{zm2}) that
$\tilde v$ is an eigenfunction of $B$ with eigenvalue zero.
By explicit computation, we see that according to (\ref{(15)}), 
(\ref{(20)}), and (\ref{(17)}),
the corresponding eigenfunction ${\bf \Psi}$ of $A$ has components in
the $j=1$ and $j=-1$ subspaces.

In a similar way, setting $u_\tau (x,y,t)\equiv u_0(x,y-\tau,t)$ yields
the eigenfunction $\tilde v=\D_y v_0$
of the operator $B$ with eigenvalue zero, and the corresponding
eigenfunction of $A$ has components in the $j=1$ and $j=-1$ subspaces. 
It is straightforward to check that these two eigenfunctions resulting
from translations along the $x$ and $y$ axes are linearly independent.

\item Invariance of NLS under change of standing-wave frequency:

As mentioned in Section \ref{tax}, localized standing-wave solutions
$u_0=e^{i\omega\,t}e^{i\,m\theta }w(r)$ of \eq{1nls} exist for a range of 
frequencies $\omega$.  The radial profile $w(r)$ depends on $\omega$
through the differential equation (\ref{1weq}), in which
$f(w)=g(w)-\omega w$.  Consequently, $v_0$ as well as $u_0$ is 
parametrized by $\omega$. Differentiating with
respect to parameter $\omega$, we have 
\be
\tilde u=itu_0 + 
e^{i\omega\,t} \D_\omega v_0, \quad 
\tilde v=itv_0 + \D_\omega v_0.
\ee
Equation (\ref{zm2}) thus gives
$ B\tilde v=iv_0$.
Since we established earlier that $B(iv_0)=0$, we see that
$ B \left(\D_\omega v_0 \right)=iv_0$.
Thus (again since $B(iv_0)=0$),
\be
B^2 \left( \D_\omega v_0\right)=0.
\ee
That is, $\D_\omega v_0$ is a
generalized eigenfunction of B of index 2 corresponding to
eigenvalue zero.
Because
\be
\D_\omega v_0 = 
e^{i\,m\theta}\D_\omega w(r),
\ee
the angular dependence of the generalized eigenfunction
$\D_\omega v_0$
is identical to that generated by invariance under phase changes, and as
a result the corresponding generalized eigenfunction $\Psi$ of $A$ lies
in the $j=0$ subspace.

\item Invariance of NLS under velocity boost:

Let $c\in\R^2$ be a constant vector.  If $u(x,t)$ is a solution to \eq{1nls},
then so is the function
\be
U_c(x,t)\equiv\exp \left\{ {i\left( {{{1 \over 2}}c\cdot x-
{{1 \over 4}}\left| c \right|^2t} \right)}\right\}
\,\,u(x-c\,t,\,\,t).
\ee
Again writing $(x,y)$ for the components of the spatial coordinate
vector, we first let $u_\tau$ be the standing wave $u_0$ boosted with
speed $\tau$ in the $x$ coordinate direction (that is, $c=(\tau,0)$):
\be
u_\tau(x,y,t)\equiv\exp \left\{ {i\left( {{{1 \over 2}}\tau \,
x-{{1 \over 4}}\tau ^2t} \right)} \right\}\,\,
u_0(x-\tau \,t,\,\,y,\,\,t).
\ee
Then  
$\tilde u={{i\over 2}}x\,u_0-
t\D_x u_0$ and
$\tilde v={{i\over 2}}x\,v_0-
t\D_x v_0$.
Equation (\ref{zm2}) thus gives
$ B\tilde v=-\D_x v_0$.
Since we established earlier that 
$B\left(\D_x v_0\right)=0$, 
we see that
$ B(ixv_0) = -2\D_x v_0$.
Therefore (again since 
$B\left(\D_x v_0\right)=0$),
\be
B^2 \left(ixv_0\right)=0.
\ee
That is, $ixv_0$ is a
generalized eigenfunction of B of index 2 corresponding to
eigenvalue zero.
By explicit computation, we see that 
the corresponding eigenfunction ${\bf \Psi}$ of $A$ has components in
the $j=1$ and $j=-1$ subspaces.

In a similar way, 
letting $u_\tau$ be the standing wave $u_0$ boosted with
speed $\tau$ in the $y$ coordinate direction (that is, $c=(0,\tau)$),
we find that $iyv_0$ is a
generalized eigenfunction of B of index 2 corresponding to
eigenvalue zero, and the corresponding
eigenfunction of $A$ has components in the $j=1$ and $j=-1$ subspaces. 
The two generalized eigenfunctions $ixv_0$ and $iyv_0$
resulting from velocity boosts in the $x$ and $y$ directions are linearly
independent.

\end{itemize}

We have thus found two $j=0$ generalized eigenfunctions, and four
$|j|=1$ generalized eigenfunctions, for eigenvalue
$\lambda=0$, as summarized in Table 2. 

\begin{table}\label{T.zeromodes}
\begin{center}
\caption{Zero modes}
\medskip
\begin{tabular}{|c|c|c|c|l|}
\hline
Transformation & $|j|$ & Index & Generalized eigenfunction \\ \hline
{Phase, rotation, time translation} & 0 & 1  & $i v_0$ \\
{Standing-wave frequency change} & 0 & 2 &${\D_\omega{v_0}}$ \\
{Space translation in } $x$ { direction} & 1 & 1 &${\D_x{v_0}}$\\ 
{Space translation in } $y$ { direction} & 1 & 1 &${\D_y{v_0}}$ \\ 
{Boost in } $x$ { direction} & 1 & 2 & $i x v_0$  \\
{Boost in } $y$ { direction} & 1 & 2 & $i y v_0$  \\
\hline \end{tabular}\end{center}
\end{table}

To establish that these Jordan chains of generalized eigenfunctions do in fact
terminate with index 2 in the general case, we make use of the following
observation.

{\lemma
Suppose $A=JL$ where $L$ is self-adjoint and $J^*=J^{-1}=-J$.
Suppose $\Psi_K$ is a generalized eigenfunction of $A$ with index $K$
corresponding to eigenvalue zero.
If there exists $\Phi$ in the domain of $A$
such that $A\Phi=\Psi_K$, then 
$\left<\Psi_j \left| J\Psi_{K+1-j}\right.\right>=0$ for $j=1,\dots,K$,
where $\Psi_j\equiv A^{K-j}\Psi_K$.
\label{endchain}}

\proof
From the definition of $\Psi_j$ it follows that $A\Psi_j=\Psi_{j-1}$ and
$A^j\Psi_j=0$ but $A^{j-1}\Psi_j\ne 0$.  Thus $\Psi_j$ is a generalized
eigenfunction of $A$ with index $j$.

Define $\Psi_j^*\equiv J\Psi_j$. From the assumed form for $A$ we then have
\be
A^*\Psi_j^* = -JA\Psi_j = -J\Psi_{j-1} = -\Psi_{j-1}^*.
\ee
It follows that
\be
(A^*)^j\Psi_j^* = (-1)^j JA\Psi_1 = 0.
\ee
Thus, for $1\le j\le K$, we have
\be
\ba{rcl}
\left< \Psi_j \left| J\Psi_{K+1-j} \right.\right>&=&
-\left< \Psi_j^* \left| \Psi_{K+1-j} \right.\right>
=-\left< \Psi_j^* \left| A^{j-1}\Psi_{K} \right.\right>
\cr
&=&-\left< \Psi_j^* \left| A^{j-1}A\Phi \right.\right>
=-\left< \Psi_j^* \left| A^j\Phi \right.\right>
\cr
&=&-\left< (A^*)^j\Psi_j^* \left| \Phi \right.\right>
=0,
\ea
\ee
as asserted.
\eproof

We now apply this observation to our index-2 generalized eigenfunctions. 
Consider first the generalized eigenfunction $\psi_2=ixv_0$ of $B$ arising from
velocity boost in the $x$ direction.  Then 
$\psi_1\equiv B\psi_2=-2\D_x v_0$ and we have
\be
\ba{rcl}
\left< \Psi_1 \left| J\Psi_2 \right.\right>
=\Re\left< \psi_1 \left| i\,\,\psi_2 \right.\right>
=\Re \left<-2\D_x v_0
\left| i\,\,ixv_0\right.\right>
=-\left\|{v_0}\right\|^2\ne 0.
\ea
\ee
Thus there is no generalized eigenfunction in this Jordan chain with index 3
(or higher).
Similarly, analyzing the index-2 generalized eigenfunction $\psi_2=iyv_0$ of $B$
arising from velocity boost in the $y$ direction, we find that there is no
generalized eigenfunction in this chain with higher index.

We next consider the index-2 generalized eigenfunction 
$\psi_2=\D_\omega v_0$
of $B$ arising from change of standing-wave frequency, which lies in the $j=0$
subspace.
We set $\psi_1\equiv B\psi_2 = iv_0$, and we find
\be
\ba{rcl}
\left< \Psi_1 \left| J\Psi_2 \right.\right>
=\Re\left< \psi_1 \left| i\,\,\psi_2 \right.\right>
=\Re \left<iv_0\left| 
i\,\,\D_\omega v_0\right.\right>
=\half\D_\omega \left\|{v_0}\right\|^2.
\ea
\ee
Generically, we expect the quantity
$\D_\omega \left\|{v_0}\right\|^2$ to be nonzero, 
hence that $\psi_2=\D_\omega v_0$
is the highest-index generalized eigenfunction in this Jordan chain.  Our
numerical studies confirm this expectation for 
nonlinearities other than pure cubic.

The six-dimensional eigenspace for eigenvalue zero summarized in the 
table is in agreement with our numerical observations.  For 
nonlinearities other than pure cubic, 
we find numerically that the eigenvalue zero has
multiplicity two in each of the $j=0$, $j=1$, and $j=-1$ subspaces.

In contrast, in the special case of a purely cubic nonlinearity 
$g(u)=\gamma|u|^2u$ in two spatial dimensions,
the norm $\left\|{v_0}\right\|^2$ is independent of $\omega.$  
To see this, we note that the function ${\tilde v}_0$ defined by 
$v_0(x)=\sqrt{{\omega/{\gamma}}} \,\,
{\tilde v}_0(\sqrt{\omega}\,\,x)$ satisfies the $\omega$-independent equation
$\Delta{\tilde v}_0-{\tilde v}_0+|{\tilde v}_0|^2{\tilde v}_0=0$.  It is
straightforward to compute that, in $N$ spatial dimensions,
$\left\|{v_0}\right\|^2={{1\over{\gamma}}}\omega^{(2-N)/2}
\left\|{\tilde v}_0\right\|^2$.
Thus, for the pure-cubic nonlinearity, in $N=2$ spatial dimensions, the index-2
generalized eigenfunction does not necessarily terminate the Jordan chain in
the $j=0$ subspace, and in fact generalized eigenfunctions
with indices 3 and 4 are present in this special case.

The structure of the generalized null spaces of $A$ in the case of \eq{1nls}
with pure-power nonlinearities is established in Appendix B of the 
paper \ccite{W85}.
The index-3 generalized eigenfunction
$\psi_3=ir^2v_0$ of $B$ arises from the invariance of cubic NLS in two
spatial dimensions under the Talanov lens transformation.
An index-4 generalized eigenfunction $\psi_4$ has the form
$\psi_4=\xi(r)e^{im\theta}$, where $\xi$ is the bounded solution to the
ordinary differential equation $L_3\xi=r^2w_0$, where 
\[
L_3={{\D^2\over\D r^2}+{1\over r}{\D\over\D r}-{m^2
\over r^2}} -\omega +3\gamma w_0^2.
\]
There are no generalized eigenfunctions with higher index.
Our numerical observations in the pure cubic case show that eigenvalue
zero has multiplicity four in the $j=0$ subspace, in agreement with this
analysis.

\section{Two proofs}
\reseteqnos

It remains to prove two technical results needed
to characterize eigenvalues.

\subsection{Proof of Lemma \ref{4yasym}}\label{Pf4.1}

To prove Lemma \ref{4yasym}, we will transform the system \eq{41yf}
\[
\yf' = \Bf(r,j,\lambda)\yf
\]
to an equivalent system of the form
$\zetaf' = (\halo{\Cf}+\tCf)\zetaf$ where $\halo{\Cf}$ is constant and 
$\tCf$ is integrable, then make use of facts concerning asymptotic behavior of
solutions to such systems.  To do so,
we make the change of variables $\zeta_\pm (x) = h(r)y_\pm (r)$, 
where $x=x(r)$ and $h(r)$ are smooth monotonic functions such that
\be \label{41xr}
x(r)=\cases{ \ln r & for $r<\shalf$, \cr r & for $r>2$,}
\quad h(r)= \cases{1 & for $r<\shalf$,\cr \sqrt{r} & for $r>2$.}
\ee
With this change of variables, system \eq{41yf} is equivalent to 
\be
\zetaf' = \Cf(x)\zetaf ,
\label{41zetaf}
\ee
where
\be\label{41M}
\zetaf(x)=\pmatrix{\zeta_+ \cr \zeta'_+ \cr \zeta_- \cr \zeta'_-} =
M(r)\yf(r), \quad
M(r) = \pmatrix{
h &0&0&0\cr
h'r_x &hr_x &0&0\cr
0&0&h&0\cr
0&0&h'r_x&hr_x} ,
\ee
and 
\be \label{41C}
\Cf(x)= r_x(M\Bf M^{-1}+M'M^{-1}).
\ee
Note that
\be \label{41M2}
\pmatrix{h&0\cr h'r_x&hr_x}
= \pmatrix{1&0\cr 0&r}, \mbox{\quad resp. \quad}
\pmatrix{\sqrt{r}&0\cr \frac{1}{2\sqrt{r}}& \sqrt{r}},
\ee
for $r<\frac12$ and $r>2$.

We can write  $\Cf(x)=\halo{\Cf}+\tCf(x)$, where
\be \label{41Chalo}
\halo{\Cf}=\pmatrix{0&1&0&0\cr
(j+m)^2&0&0&0\cr 0&0&0&1\cr 0&0&(j-m)^2&0},
\quad\mbox{resp.}\quad 
\pmatrix{0&1&0&0\cr
k_+^2&0&0&0\cr 0&0&0&1\cr 0&0&k_-^2&0},
\ee
for $x<-\ln 2$ and $x>2$,
and where the entries of $\tCf(x)$ are 
$O(1/x^2)$ as $x\to\infty$ and $O(e^{2x})$ as $x\to-\infty$, 
so that $\tCf(x)$ (but not $x\tCf(x)$) is absolutely integrable on $\R$.

We discuss the asymptotic behavior of solutions as $|x|\to\infty$ in the
intervals $(2,\infty)$ and $(-\infty,-\ln 2)$ separately.  
We note first that because $k_+$ and $k_-$ are
nonzero, the matrix $\halo{\Cf}$ is constant and diagonalizable on
$(2,\infty)$, with eigenvalues $k_+$, $k_-$, $-k_+$, and $-k_-$.
The system $\zetaf'=\halo{\Cf}\zetaf$ on $(2,\infty)$ has the fundamental set of
solutions
\be \label{4hCset}
e^{k_+ x}\pmatrix{1\cr k_+\cr 0\cr 0\cr},\quad 
e^{k_- x}\pmatrix{0\cr 0\cr 1\cr k_-\cr},\quad
e^{-k_+ x}\pmatrix{1\cr -k_+\cr 0\cr 0\cr},\quad
e^{-k_- x}\pmatrix{0\cr 0\cr 1\cr -k_-\cr}.
\ee
It follows
from  Theorem IV.1 of \ccite{Coppel} that for each member of 
this fundamental set, 
there exists a solution $\zetaf$ of \eq{41zetaf} that is asymptotic to
it as $x\to\infty$.
Undoing the change of variables on the interval $(2,\infty)$, we see
that system \eq{41yf} has corresponding solutions respectively 
asymptotic to 
\be
\frac{e^{k_+ r}}{\sqrt{r}} \pmatrix{1 \cr k_+\cr0\cr0},\quad
\frac{e^{k_- r}}{\sqrt{r}} \pmatrix{0\cr0\cr 1 \cr k_-},\quad
\frac{e^{-k_+ r}}{\sqrt{r}} \pmatrix{1 \cr -k_+\cr0\cr0},\quad
\frac{e^{-k_- r}}{\sqrt{r}} \pmatrix{0\cr0\cr 1 \cr -k_-}
\ee 
as $r\to\infty$.
Since the modified Bessel functions $I_\nu$ and $K_\nu$ have
the asymptotic forms 
\be \label{4IKinf}
\ds I_{\nu}(z) = \frac{1}{\sqrt{2\pi z}}\,e^{z}
\left(1+O\left(\frac{1}{|z|}\right)\right),
\quad  \ds K_{\nu}(z) = \sqrt{\frac{\pi}{2z}}\,e^{-z}
\left(1+O\left({\frac{1}{|z|}}\right)\right)
\ee 
as $|z|\to\infty$ in the right half complex plane, 
we see that the
system \eq{41yf} has solutions asymptotic to $\hyf_i$ as
$r\to\infty$, for each $i=1,\ldots,4$ as claimed.

To analyze the asymptotic behavior of solutions as $r\to 0^+$, we 
first suppose
that both $j+m$ and $j-m$ are nonzero. Then
the matrix $\halo{\Cf}$ is constant and diagonalizable on
$(-\infty,-\ln 2)$, with eigenvalues $|j+m|$, $|j-m|$, $-|j+m|$, and 
$-|j-m|$. The system  $\zetaf'=\halo{\Cf}\zetaf$ on $(-\infty,-\ln 2)$
has the fundamental set of solutions
\be
e^{|j+m| x}\pmatrix{1\cr \scriptstyle |j+m|\cr 0\cr 0\cr},\ %
e^{|j-m| x}\pmatrix{0\cr 0\cr 1\cr \scriptstyle |j-m|\cr},\ %
e^{-|j+m| x}\pmatrix{1\cr  \scriptstyle-|j+m|\cr 0\cr 0\cr},\ %
e^{-|j-m| x}\pmatrix{0\cr 0\cr 1\cr  \scriptstyle-|j-m|\cr}.
\ee
It follows as before
that for each of these,
there is a solution $\zetaf$ of \eq{41zetaf} asymptotic to it as $x\to-\infty$.
Undoing the change of variables on the interval $(-\infty,-\ln 2)$, we
see that the  system \eq{41yf} has corresponding solutions 
respectively asymptotic to
\be
r^{|j+m|} \pmatrix{1\cr \frac{|j+m|}{r}\cr 0\cr0},\ %
r^{|j-m|} \pmatrix{0\cr0\cr1\cr \frac{|j-m|}{r}},\ %
r^{-|j+m|} \pmatrix{1\cr -\frac{|j+m|}{r}\cr 0\cr0},\ %
r^{-|j-m|} \pmatrix{0\cr0\cr1\cr -\frac{|j-m|}{r}}
\ee
as $r\to 0^+$.
Since for $\nu>0$ the modified Bessel functions $I_\nu$ and $K_\nu$ have
the asymptotic forms 
\be \label{4IK0}
\ds I_{\nu}(z) = \frac{(\shalf z)^\nu}{\Gamma(\nu+1)}(1+O(|z|)),\quad  
\ds K_{\nu}(z) = \frac{\shalf\Gamma(\nu)}{(\shalf z)^\nu}(1+O(|z|)) 
\ee 
as $z\to 0$, 
we see that the  system \eq{41yf} has a solution asymptotic to 
$\hyf_i$ as $r\to0$ for each $i=1,\ldots,4$, as claimed.

If either $j+m=0$ or $j-m=0$, then on
$(-\infty,-\ln 2)$ the matrix $\halo{\Cf}$ is constant and has
Jordan canonical form with a block of order $2$ corresponding to
the eigenvalue zero. Suppose for definiteness that $j=m>0$.  Then on
$(-\infty,-\ln 2)$ we have
\be
\halo{\Cf}=\pmatrix{0&1&0&0\cr 4m^2&0&0&0\cr 0&0&0&1\cr 0&0&0&0} ,
\ee
and the system $\zetaf'=\halo{\Cf}\zetaf$ has the fundamental set of
solutions
\be
e^{2mx}\pmatrix{1\cr 2m\cr 0\cr 0\cr},\quad
\pmatrix{0\cr 0\cr 1\cr 0\cr},\quad
e^{-2mx}\pmatrix{1\cr -2m\cr 0\cr 0\cr},\quad
\pmatrix{0\cr 0\cr x\cr 1\cr}.
\ee
Since the
entries of the matrix $\tCf(x)$  decay exponentially to zero as 
$x\to -\infty$, $x \tCf(x)$ is absolutely integrable on 
$(-\infty,-\ln 2)$, and we may apply Theorem IV.4 of \ccite{Coppel} to
establish that for each member of this set of solutions, 
there exists a solution $\zetaf$ of \eq{41zetaf} that is asymptotic to
it as $x\to-\infty$.
Undoing the change of variables on the interval $(-\infty,-\ln 2)$, we
see that the  system \eq{41yf} has corresponding solutions 
respectively asymptotic to 
\be
r^{2m}\pmatrix{1\cr {2m}/{r}\cr0\cr0},\quad
\pmatrix{0\cr0\cr1\cr0},\quad
r^{-2m}\pmatrix{1\cr -{2m}/{r}\cr0\cr0},\quad
\pmatrix{0\cr0\cr \ln r\cr 1/r}
\ee
as $r\to0$.
Since $I_0(z)\sim 1$ and $K_0(z)\sim -\ln z$ as $z\to 0$,
we see that the  system \eq{41yf} in this case again
has a solution asymptotic to 
$\hyf_i$ as $r\to0$ for each $i=1,\ldots,4$. This finishes the proof
of the Lemma.
\eproof

\subsection{Proof of Proposition \ref{4Eana}}\label{PfAna}

It is evident that for any given $\lambda\in\Ccut$,
the exterior systems associated with \eq{41yf} and its 
adjoint \eq{4zf} have the respective solutions 
$\ysix=\ylwyl$ and $\zsix=\zrwzr$ with the desired 
asymptotic behavior, since 
\be \label{45asy}
\ba{rcl}
\ylwyl &\sim&   
\hylwyl \quad\mbox{as $r\to0^+$,}\cr
\zrwzr&\sim&   
\hzrwzr \quad\mbox{as $r\to\infty$.}\cr
\ea
\ee
It remains to establish the uniqueness and analyticity of these solutions.
Our plan is to transform the associated exterior systems to corresponding
systems on the whole line by changing variables as in the proof of Lemma \ref{4yasym}. 
Under this transformation, the quantities $\ylwyl$ and $\zrwzr$ map to solutions
having asymptotic normalizations that enjoy a maximal rate of decay property.
Uniqueness and analyticity of these solutions is a consequence of the theory of \ccite{PW92}.

Let $x(r)$, $h(r)$, and $M(r)$ be as in \eq{41xr} and \eq{41M}.  
Suppose $\ysix(r)$ is a solution of
the exterior system
\be \label{45ysix}
\ysix' = \Bfsix(r)\ysix
\ee
associated with \eq{41yf}. Then $\ysix$ may be written as a linear combination
of solutions of the form $\yf\wedge\tyf$ where $\yf$ and $\tyf$ are solutions
of \eq{41yf}. 
(Do this first for initial data, then solve the initial-value problem.)
Given such solutions, 
then $\zetaf(x):=M(r)\yf(r)$ and $\tzetaf(x):=M(r)\tyf(r)$ 
are solutions of \eq{41zetaf}, and $\zetas:=\zetaf\wedge\tzetaf$ is a solution
of the exterior system
\be \label{45zetas}
\zetas' = \Cfsix(x) \zetas
\ee
associated with \eq{41zetaf}. Note that 
\[
\zetaf\wedge\tzetaf = (M\yf)\wedge(M\tyf) = \Msix (\ywedgety),
\]
where the $6\times6$ matrix $\Msix$ has columns determined by 
$\Msix(\bme_j\wedge\bme_k)=(M\bme_j)\wedge(M\bme_k).$
It follows that if we let $\zetas(x)=\Msix(r)\ysix(r)$, then $\zetas$ 
is a solution of \eq{45zetas}.  Similarly, the converse holds: if
$\zetas$ is a solution of \eq{45zetas} and $\ysix(r)=\Msix(r)^{-1}\zetas(x)$,
then $\ysix$ is a solution of \eq{45ysix}. 
(Note that if $N=M^{-1}$, then $\hat N\Msix(\ywedgety) = \ywedgety$
for all 4-vectors $\yf$ and $\tyf$, hence $\hat N=\Msix^{-1}$.)

We transform the adjoint exterior system in similar fashion. Suppose
$\etas(x)=\zsix(r)\Msix(r)^{-1}$. Then $\zsix$ is a solution of the exterior
system
\be \label{45zsix}
\zsix' = -\zsix \Bfsix(r)
\ee
associated with \eq{4zf} if and only if $\etas$ is a solution of the exterior system
\be \label{45etas}
\etas ' = -\etas \Cfsix(x)
\ee
associated with the equation adjoint to \eq{41zetaf}.

Let us now study solutions of \eq{45etas} that correspond to solutions
of \eq{45zsix} with the asymptotic behavior in \eq{45asy}.
Using the asymptotic relations established in Lemma~\ref{4zasym},
the definitions \eq{4haloZ} and \eq{IKdef}, 
and the asymptotics in \eq{4IKinf}, we determine that 
\be \label{45esrt}
(\zrwzr)\Msix^{-1} \sim \left(\halobz_1 M^{-1}\right) \wedge \left(\halobz_2 M^{-1}\right)
\sim \exp(-\musix^\rt x) \wsixrt
\ee
as $x\to\infty$,
where $\musix^\rt=k_++k_-$ is an eigenvalue of  ${\halo{\Cf}}_\wedge$ for $x>2$,
and $\wsixrt$ is 
the particular
associated left eigenvector 
\be \label{wsix}
\wsixrt = \frac{\pi}{2\sqrt{k_+k_-}}
(k_+,1,0,0)\wedge(0,0,k_-,1).
\ee
Since $k_+$ and $k_-$ have positive real parts for $\lambda\in\Ccut$, 
it follows that
$\musix^\rt$ is a simple eigenvalue, and 
$\musix^\rt$
is the {\it unique} eigenvalue 
of largest real part.
For $\lambda\in\Ccut$, it is clear that $\musix^\rt$ and $\wsixrt$ are analytic
and the latter is nonvanishing. 

Since $\tCf_\wedge(x)=\Cf_\wedge(x)-{\halo{\Cf}}_\wedge$ is 
integrable on $\R$, we may now invoke the theory of \ccite{PW92}. 
Proposition 1.2 of \ccite{PW92} asserts (after space reversal)
that \eq{45etas} has a unique solution $\etas^\rt(x,\lambda)$ satisfying
\be \label{45etaasy}
\exp(\musix^\rt x)\etas^\rt(x,\lambda) \to \wsixrt \quad\mbox{as $x\to\infty$},
\ee
and this solution is globally analytic in $\lambda$ for $\lambda\in\Ccut$. 
It follows that $\zsix^\rt(r,\lambda):=\etas^\rt(x,\lambda)\Msix(r)$ yields the unique solution
of \eq{45zsix} having the asymptotic behavior asserted in Proposition~\ref{4Eana},
and that $\zsix^\rt$ is analytic in $\lambda$ for $\lambda\in\Ccut$. 

The treatment of \eq{45ysix} is similar, at least in the case when
$|j+m|$ and $|j-m|$ are both nonzero, which we consider first. 
As $x\to-\infty$ we have 
\be \label{45zslt}
\Msix(\ylwyl) \sim \left(M \halo{\yf_1} \right) \wedge\left(M \halo{\yf_2} \right) 
\sim \exp(\musix^\li x) \vsixlt
\ee
where $\musix^\li=|j+m|+|j-m|$ is an eigenvalue
of ${\halo{\Cf}}_\wedge$ for $x<-\ln 2$,  and
$\vsixlt$ is 
the particular
associated right eigenvector 
\be \label{vsix}
\vsixlt = \frac{ (\half k_+)^{|j+m|}(\half k_-)^{|j-m|}}
{\Gamma(|j+m|)\Gamma(|j-m|)} 
\pmatrix{1\cr |j+m|\cr 0\cr0} \wedge \pmatrix{0\cr0\cr1\cr|j-m|}.
\ee
Clearly $\musix^\li$ is independent of $\lambda$, and $\vsixlt$
is analytic and nonvanishing in the cut plane $\Ccut$. 
Provided $|j+m|$ and $|j-m|$ are nonzero, $\musix^\li$ is simple
and is the {\it unique} eigenvalue of largest real part for 
${\halo{\Cf}}_\wedge$ for $x<-\ln 2$. 
Proposition 1.2 of \ccite{PW92} implies
that \eq{45zetas} has a unique solution $\zetas^\li(x,\lambda)$
satisfying
\be \label{45zetaasy}
\exp(-\musix^\li x)\zetas^\li(x,\lambda)\to \vsixlt \quad\mbox{as $x\to-\infty$},
\ee
and this solution is globally analytic for $\lambda\in\Ccut$. 
It follows that $\ysix^\lt := \Msix^{-1}\zetas^\li$ is the unique solution
of \eq{45ysix} having the asymptotic behavior asserted in Proposition~\ref{4Eana},
and that $\ysix^\lt$ is analytic in $\lambda$ for $\lambda\in\Ccut$.

It remains to consider the case when either $|j+m|$ or $|j-m|$ is zero.
For definiteness, suppose $j=m>0$. We still have \eq{45zslt}, 
and $\musix^\li$ is the unique eigenvalue of largest real part, but now
it is a double eigenvalue of geometric multiplicity one, with
eigenspace spanned by $\vsixlt$. The proof of Proposition 1.2 of \ccite{PW92}
must be modified to obtain the uniqueness and analyticity of $\zetas^\li$.

This can be done as follows. First, we change variables in \eq{45zetas}, writing
\be \label{4vzeq}
\vz(x)= \exp(-\musix^\li x)\zetas(x)
\ee
Let $B={\halo{\Cf}}_\wedge-\musix^\li I$ and $R(x,\lambda)=\tCf_\wedge(x)$.
Then the first equation in \eq{45zetas} is equivalent to 
\[
\vz'(x) = \left(B+R(x,\lambda)\right)\vz .
\]
Note that there is a constant $C_1>0$
independent of $\lambda$ such that 
\[
\|\exp(Bx)\| \le C_1(1+|x|)
\]
for all $x\ge0$.  Also, the entries of
$R(x,\lambda)$ decay exponentially as $x\to-\infty$, uniformly for 
$\lambda$ in any compact subset of $\Ccut$. 
For any compact subset $\Omega_1\subset\Ccut$, there exists $x_0$ sufficiently
negative so that 
\[
\theta = C_1\sup_{\lambda\in\Omega_1} \int_{-\infty}^{x_0}
(1+|s|)\|R(s,\lambda)\|\,ds <1.
\]
We define an operator $\calF=\calF(\lambda)$ on the space
$C((-\infty,x_0])$ of bounded continuous functions on $(-\infty,x_0]$ by
\[
\calF \vz(x)= \int_{-\infty}^x e^{(x-s)B}R(s,\lambda)v(s)\,ds.
\]
Then it follows that for $\lambda\in\Omega_1$,
\[
\sup_{x<x_0} |\calF\vz(x)| \le \theta \sup_{x<x_0}|\vz(x)| .
\]

From here, the proof goes the same as the proof of Proposition 1.2 of \ccite{PW92}.
The equation $\vz = \tvz +\calF\vz$ yields a bicontinuous correspondence
between bounded solutions of \eq{4vzeq} on $(-\infty,x_0]$, and those of
$\tvz' = B\tvz$, all of which have
the form $\tvz(x)=c\vsixlt$ for some $c\in\C$. The solution we seek has the 
representation
\[
\zetas^\li(x) = \exp(\musix^\li x)(I-\calF)^{-1} \vsixlt
\]
for $x<x_0$, and is extended by solving \eq{45zetas}. 
It is independent of $x_0$ and is obtained by iterations that are analytic 
in $\lambda$ and converge uniformly for $x<x_0$, $\lambda\in\Omega_1$.
It follows that $\zetas^\li(x)$ is globally analytic for $\lambda\in\Ccut$.
This finishes the proof of the Proposition.
\eproof



\section*{Acknowledgements}
This material is based upon work supported by the 
National Science Foundation under Grant 
Nos.\  DMS97-04924 and DMS00-72609 and SCREMS grant DMS96-28467.
This work was partially supported by University of North Texas Faculty
Research Grants and by the University of Maryland Graduate Research
Board.  The authors thank B. A. Malomed and I. Towers for communicating 
their work \cite{TB2} prior to publication.

\bibliographystyle{plain}
\bibliography{ls}
\pagebreak

\nwc{\xsz}{4.5in}
\nwc{\ysz}{3.5in}
\nwc{\xszz}{3.5in}

\end{document}